\RequirePackage{plautopatch}
\documentclass[a4paper,11pt]{article}

\usepackage{ulem} 

\usepackage{geometry}
\renewcommand{\baselinestretch}{1.4} 

\usepackage{here}

\usepackage{multicol}
\usepackage{multirow}
\usepackage{fancybox, ascmac}

\usepackage[authoryear]{natbib}
\usepackage[hyphens]{url}
\usepackage{hhline}

\usepackage{amsmath, amssymb} 
\usepackage{bm} 
\usepackage{array, booktabs} 
\usepackage[utf8]{inputenc}
\usepackage{txfonts}
\usepackage[T1]{fontenc}

\usepackage{palatino}

\usepackage[pdftex]{graphicx}
\usepackage{tikz}
\usepackage[framemethod=tikz]{mdframed}
\usepackage{hyperref}
\hypersetup{
    colorlinks = true,
    linkcolor  = blue,
    citecolor  = blue,
    filecolor  = purple,      
    urlcolor   = blue,
    }

\usepackage{tabularx} 
\usepackage{colortbl} 
\usepackage{float} 

\def\diamondleaders{\par\vskip.5\baselineskip
 \leavevmode\hbox{}\hskip1zw\leaders
 \hbox to.5zw{\hss\footnotesize ◇\hss}
 \hfill\hskip1zw\hbox{}\par}

\makeatletter
\def\thanks#1{%
   \footnotemark
   \edef\@tempa{\noexpand\noexpand\noexpand\footnotetext[\the\c@footnote]}%
   \toks@\expandafter{\@thanks}%
   \toks\tw@{{#1}}
   \xdef\@thanks{\the\toks@\@tempa\the\toks\tw@}}
\makeatother

\begin{document}


\title{Childcare, Time Allocation, and the Life cycle: \\ Estimation for Japanese Women\thanks{We would like to express our gratitude to the discussant, Emiko Usui, and the chairperson, Daiji Kawaguchi, as well as to the participants in the 2023 Autumn Meeting of the Japanese Economic Association. We are thankful for the comments provided by Munetomo Ando, Yasusada Murata, and the participants in the seminar at Nihon University. Any remaining errors are the responsibility of the authors.  Iiboshi was supported by JSPS KAKENHI Grant Number 21K01464.} }
\date{\today}
\author{{\large Hirokuni IIBOSHI
\thanks{Coresponding author; College of Economics, Nihon University 
E-mail: \url{iiboshi.hirokuni@nihon-u.ac.jp}}\quad  \quad 
Daikuke OZAKI
\thanks{Tokyo Metropolitan University
E-mail: \url{ozaki-daisuke@ed.tmu.ac.jp}}\quad  \quad 
Yui YOSHII
\thanks{Japan International Cooperation Agency
E-mail: \url{b.kp.milov.e.y721@icloud.com}}
\thanks{Iiboshi is grateful for financial support from the Japan Society for the Promotion of Science 21K01464.}}}

\maketitle

\vspace{1\baselineskip}

\begin{abstract}
\begin{small}
\noindent

This study explores the impact of gender differences in preferences and productivity in home production on the time allocation in married couples, particularly in relation to childcare responsibilities. Using aggregated data from Japan, we estimate a life-cycle model that tracks the development of a child from infancy to adulthood by extending the work of \cite{blundellChildrenTimeAllocation2018}. Our findings reveal a decrease in maternal earnings following childbirth, aligning with the empirical evidence of the ``child penalty.'' However, the model's projections and the actual data diverge significantly during the phase of maternal earnings recovery, showing a discrepancy of approximately 50\%, the size of which implies an involuntary reduction in the wife's market work earnings.

\vspace{0.5\baselineskip}

\vspace{0.5\baselineskip}

\end{small}
\end{abstract}
\newpage









\section{Introduction}
\label{sec:INTRO}
In the 21st century, women's labor force participation rates have approached those of men, not only in Western countries but also in Asian regions, including Japan \citep{olivetti2016evolution}. Nevertheless, the enduring gender wage gap remains substantial. In recent times, there has been a notable increase in academic research focused on understanding the factors that contribute to this disparity \citep{cortes2023}. In particular, the ``child penalty,'' which predominantly affects women, has emerged as a key determinant of this gap in developed countries. This phenomenon refers to women's earnings remaining stagnant after childbirth and taking longer to recover than men's, unlike those of their non-childbearing counterparts \citep{Kleven2019b, Kleven2019a}. This effect has also been observed in Japan \citep{Kleven2023}.

Numerous studies have highlighted that this gap is due to a significant difference in time allocation between men and women. That is, in addition to paid market work, women devote considerably more time than men to unpaid household production. This disparity contributes notably to the gender gap in the labor market. The examination of gender-specific time allocation dates back to the foundational work of \citet{Becker1965} and \citet{Becker1975} in family economics. They emphasized the impact of childbirth and childcare responsibilities on parents' allocation of time and resources. Recent research, such as that by \citet{cortes2023}, has shed light on the relationship between the gender gap in labor market outcomes and home production, particularly with respect to gender differences in preferences and productivity in housework or childcare. These revelations revolve around parental time allocation. Moreover, researchers have often noted the complex relationship between maternal career progression and childcare, which can pose challenges for working mothers.

This study investigates how childcare affects the time allocation of men and women throughout the life cycle, using data from Japan, where there is an extremely large difference in the amount of time spent on childcare between couples.\footnote{According to the Survey on Time Use and Leisure Activities, Japanese women aged 20--30 spend approximately four to five times more of their own time than men on housework and childcare.}
This context is significant due to the increasing number of working mothers.

To achieve this, we adapt the unitary model developed by \citet{blundellChildrenTimeAllocation2018}, who analyzed couples in the United States. We apply their model to Japanese aggregate data and extend it by incorporating a life-cycle model that accounts for the child's age, spanning from birth to adulthood.\footnote{As will be shown in the following sections, the life-cycle model is becoming an important framework for analyzing women's labor supply. Furthermore, centering the analysis on the timing of childbirth, some research, such as that by \cite{Attanasio2008}, has employed the life-cycle model to examine women's labor supply. \cite{adda2017career_cost} also used the life-cycle model to examine women's choice of occupation as well as their fertility. However, their life-cycle model only includes consumption and labor participation, omitting the aspect of time allocation.}
This extension is crucial for a comprehensive understanding of how married couples allocate their time between market work, leisure, and housework, including parenting, throughout their journey as parents.

Prior research, including the work of \citet{blundellChildrenTimeAllocation2018}, has not explicitly addressed the influence of a child's age on parental time allocation. We use a life-cycle model to provide a theoretical foundation for explaining the mechanisms through which differences in childcare burdens at different stages of child development affect parents' time allocation. Furthermore, this study seeks to evaluate how parental income inequality, based on educational attainment, affects maternal time allocation.

Our research adopts a quantitative approach to policy analysis. Accordingly, following \cite{Greenwood2016}, we estimate parameters from the generalized method of moments (GMM) and use the Survey on Time Use and Leisure Activities (STULA), a dataset from a survey involving approximately 200,000 respondents in Japan. With a model calibrated to this dataset, we conduct policy simulations to assess quantitatively the impact of income benefit for parental leave  (PL) on parenting hours as well as the effect of nursery school utilization on parental working hours and leisure hours.

Our estimation results reveal valuable insights. Our model's outcomes align with the 85\% reduction in maternal earnings observed in the context of the child penalty phenomenon after childbirth. However, notable disparities between our model's calculated results and the empirical data arise during the subsequent phase of maternal earnings recovery.
This incongruity becomes prominent about three years post-childbirth, with a peak deviation of around 50\%. This suggests that the extent of earnings reduction linked to the child penalty is a consequence of the mothers' constrained participation in the labor market. In particular, up to the second year after childbirth, mothers devote more time to childcare. However, the results also suggest that the decline in maternal earnings after the third year post-childbirth cannot be attributed solely to reduced working hours to accommodate childcare. Factors that may contribute to the persistence of these deviations include mothers' transition to jobs with flexible working hours but lower wage rates and a wage structure that relies on experience and tenure.

Furthermore, our estimated results indicate that couples with a college education or higher are more affected by policy changes than those with less than a high school education. Amplified lifetime earnings and accumulated assets result in a 6\% increase in women's utilization of PL and a reduction of approximately 30 minutes in daily working hours. In contrast, using nursery schools increases women's working hours by an average of about two hours, largely due to reduced childcare responsibilities. However, this adoption also leads to about a 1.2\% decline in women's utilization of PL. Notably, a 25\% increase in the income replacement rate (RR) for PL corresponds to an almost 20\% rise in its utilization. Similarly, a consistent 10\% increase in wages results in a 3.0\% increase in PL. Moreover, higher family assets are linked to reduced female working hours and increased childcare hours.

The paper's structure is as follows. In Section \ref{sec:Lit} we will briefly conduct a literature review. Section \ref{sec:DATA} provides a summary of the aggregated time allocation data for Japan and presents statistical evidence derived from data regression. In Section \ref{sec:MODEL}, the model employed in this study is elaborated. Section \ref{sec:PARAMETER_SET} delineates the methodology and parameter configurations adopted in the numerical computations and estimation. The findings of the numerical results are presented in Section \ref{sec:NUMEARICAL}, followed by the policy simulation, which is described in Section \ref{sec:SIMULATION}. Lastly, Section \ref{sec:CONCLUSION} encapsulates the conclusions drawn from the study.

\section{Literature Reviews}
\label{sec:Lit}

In recent years, numerous studies have been conducted on women's labor supply and time allocation, as well as the gender gap, and the effects of related policies have been examined. \citet{Albanesi2023} and \citet{Bertrand2020} provided comprehensive surveys of these topics. \citet{Greenwood2018} also explored these subjects in their textbook. In the context of the gender gap in time allocation, the U.S. was covered in the studies by \citet{Aguiar2007}, \citet{Aguiar2013}, and \citet{Ramey2010}, while \citet{Gimenez-Nadal2023} provided a comparison of major Western industrialized countries. \citet{Albanesi2009} additionally offered a theoretical analysis of the time allocation between women's home production and their labor force participation. Furthermore, \citet{Albanesi2016} are among those who attributed the increase in women's labor force participation to technological progress in healthcare and childcare. Similarly, \citet{Greenwood2016, Greenwood2023} examined family structures and the nature of women's labor through the lens of long-term technological changes. For a theoretical exploration of women's roles in society and the influence of social norms, \citet{Bertrand2021}'s work is recommended. For an analytical perspective on the female labor supply using the life-cycle model, refer to \citet{Blundell2016a, Blundell2016b, Blundell2021}. 

Relocating our focus to Japan, a country emblematic of the global trends discussed earlier, we can observe a burgeoning body of research in recent years dedicated to the economic analysis of women's household chores and their participation in the labor market. In contrast to many of these empirical studies, our research analyzes the impact of childbearing and childcare on parents' choices at each stage of their lives using a life-cycle model.  Among these studies, \citet{Kitao2023} employed the same statistics as ours to scrutinize quantitatively how technological progress in household chores has influenced Japanese women's time allocation patterns over time. \citet{Yamaguchi2019} quantitatively analyzed the impact of acquiring PL on women's labor supply using panel data on women. The study by \citet{Kawaguchi2022} underscored that women's earnings do not reflect their skills when compared with men's earnings, revealing the existence of a gender wage disparity. \citet{Usui2016} investigated how the division of household responsibilities between spouses influences women's job satisfaction, while \citet{Usui2023} delved into the repercussions for the gender gap of heightened parental expectations for sons compared with daughters. \citet{Kawaguchi2009} reported that sons raised by working mothers are more inclined to support their wives in pursuing gainful employment. Finally, The study by \citet{Usui2017} demonstrated how the work dynamics between wives and husbands can influence the duration of breastfeeding.

\section{Statistical Evidence from Aggregate Data}
\label{sec:DATA}

\subsection{Time Allocation during the Last Decade}

This study uses the STULA data on time allocation in Japan, which are obtained through a comprehensive survey conducted every 5 years from 2011 to 2021, involving approximately 200,000 individuals, by the Ministry of Internal Affairs and Communications. This dataset captures their daily time allocation patterns and participation in various leisure activities, such as studying, self-development, volunteering, sports, hobbies, and travel. This survey also sheds light on the balance between work and personal life, making it a key measure for achieving gender equality.

Our investigation categorizes daily time into four distinct groups: (1) working time, which includes activities like commuting and work; (2) childcare time, involving caregiving and childcare duties; (3) household production time, encompassing tasks like housework; and (4) leisure time, covering periods outside of work and household tasks. These allocations are further analyzed based on gender, age, and life stages, including scenarios like being single, being married without children, and families with children of varying ages, from infancy to over 18 years.

\begin{figure}
\caption{Data: Time Allocation}
\label{fig:data_time_allocation}
\begin{center}

(a) Mothers

\includegraphics[width=16cm,height=10cm]{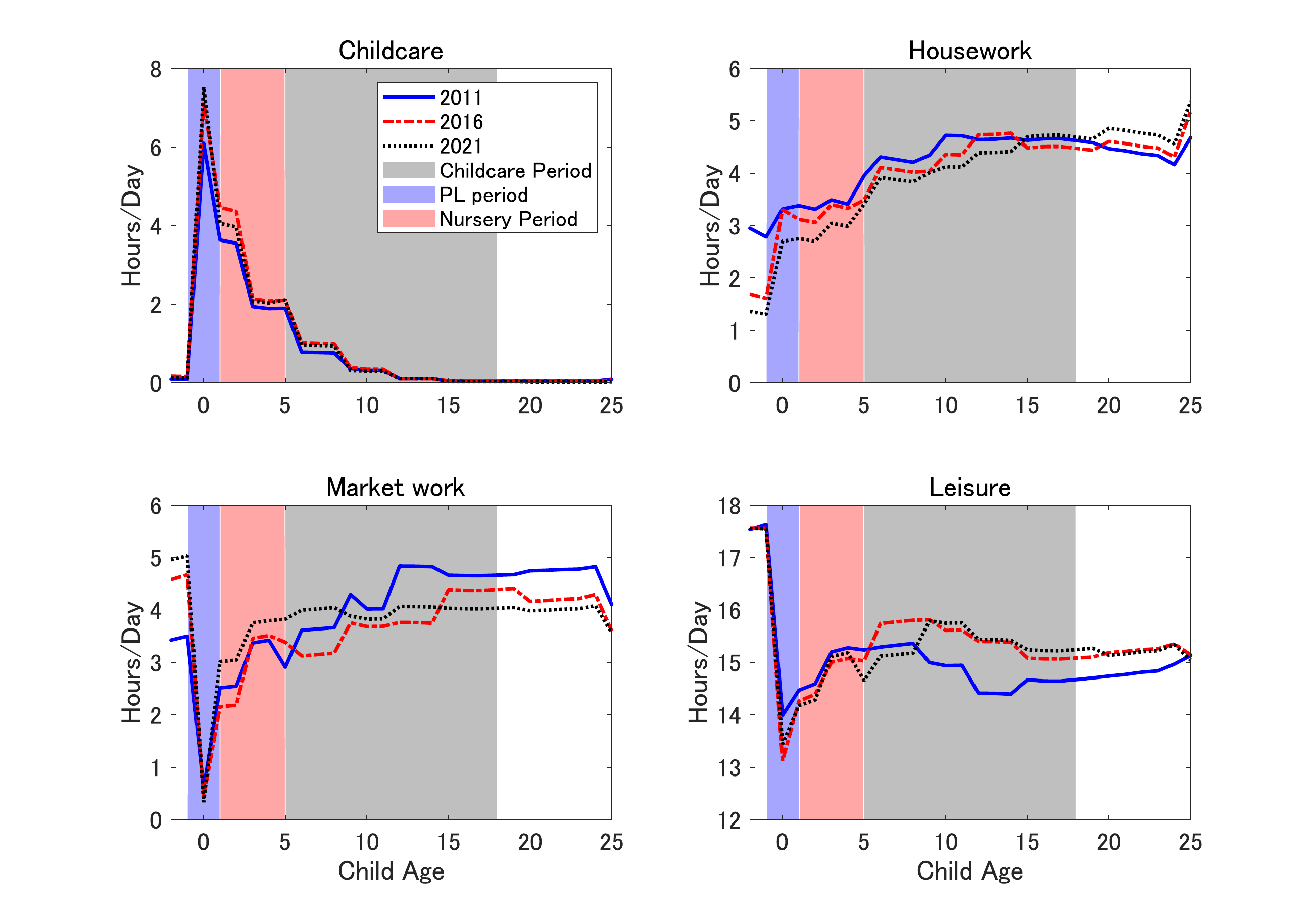}

\bigskip{}

(b) Fathers

\includegraphics[width=16cm,height=10cm]{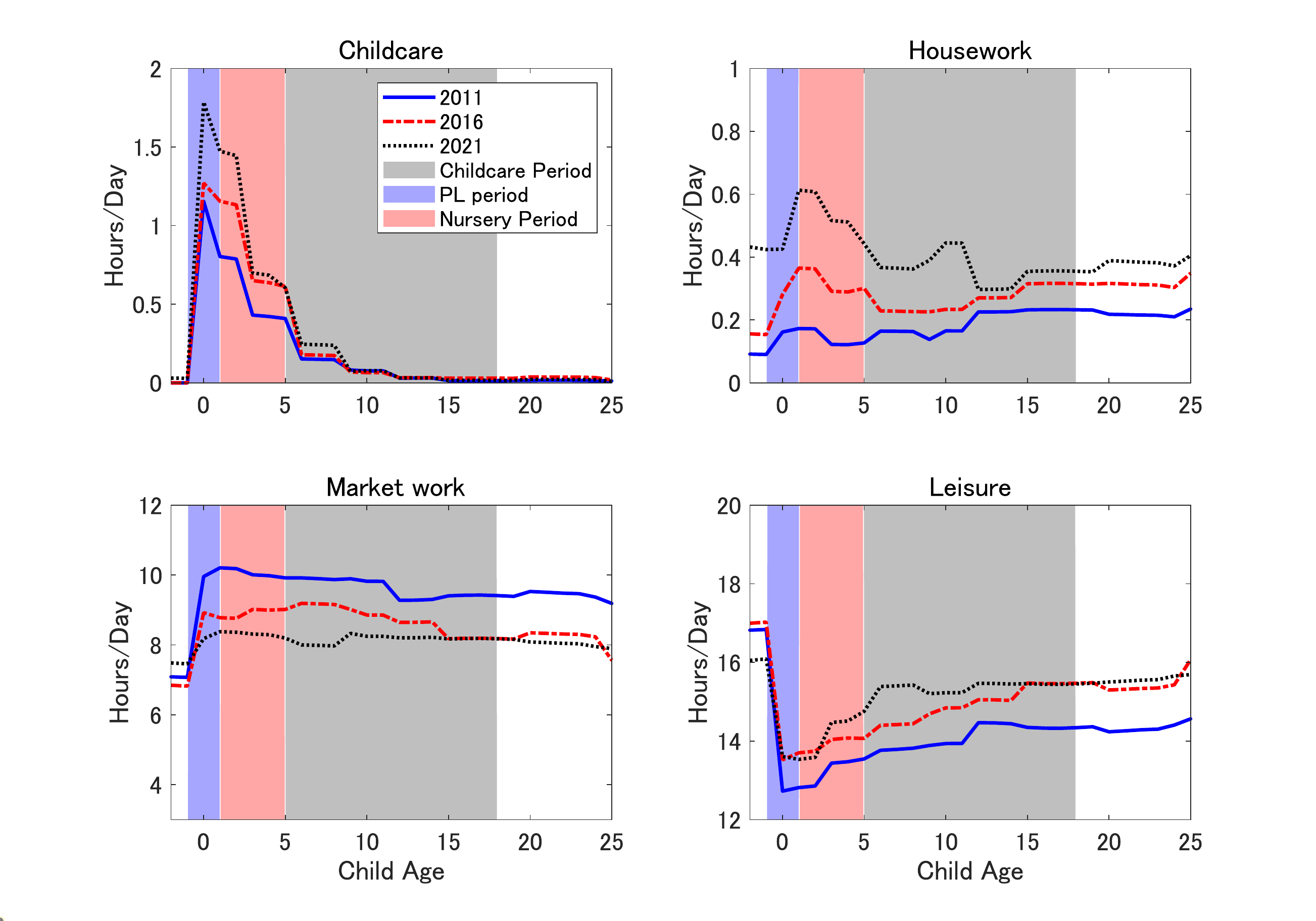}

\end{center}
\end{figure}

Figure \ref{fig:data_time_allocation} offers a comprehensive overview of the survey findings. Panel (a) presents the time allocation patterns for mothers, while panel (b) depicts those for fathers. The horizontal axes of the graphs display data aggregated by the youngest child's age, categorized by color codes representing different stages of child development. The light blue shade represents the PL phase, followed by the light red shade illustrating the nursery period, and finally, the light gray shade indicates the period up to the age of 18 when a child requires support. Panel (a) of Figure 1 reveals a clear correlation between parenting time and children's age for mothers. Immediately after childbirth, parents devote around 6--7 hours daily to childcare, resulting in reduced market work and leisure activities. These trends remain fairly stable from 2011 to 2016. When a child is younger, the childcare time gradually increases over time, while the housework hours decline. In Panel (b), similar trends emerge for fathers, with childcare time aligned with child age.

However, fathers spend significantly less time on childcare than mothers, usually around one to two hours. They also spend less time on housework, about 30 minutes. However, men's time spent on childcare and housework gradually increases from 2011 to 2021. Market working hours have decreased by nearly two hours since the last survey. Notably, the durations of market work and leisure remain consistent regardless of the child's age.

\subsection{Additional Evidence by Regression} 

\begin{figure}
\caption{Regression: Time Allocation}
\label{fig:regression_time_allocation}
\begin{center}
\begin{small}
    
\vspace{-0.7\baselineskip}

(a) Age of the Youngest Child

\includegraphics[width=148mm, height=73mm]{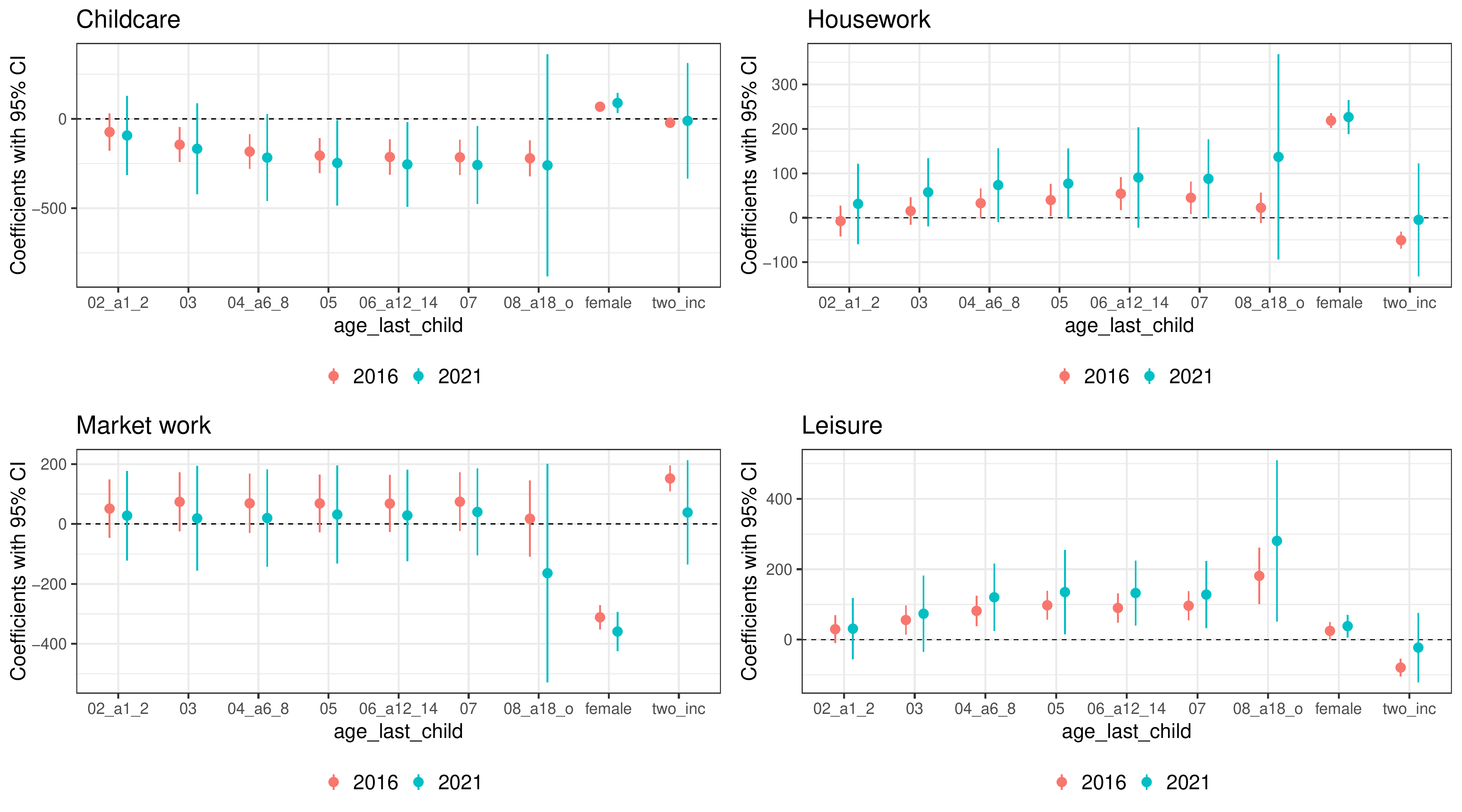}

\vspace{-0.2\baselineskip}

(b) Time Spent at Nursery

\includegraphics[width=138mm, height=73mm]{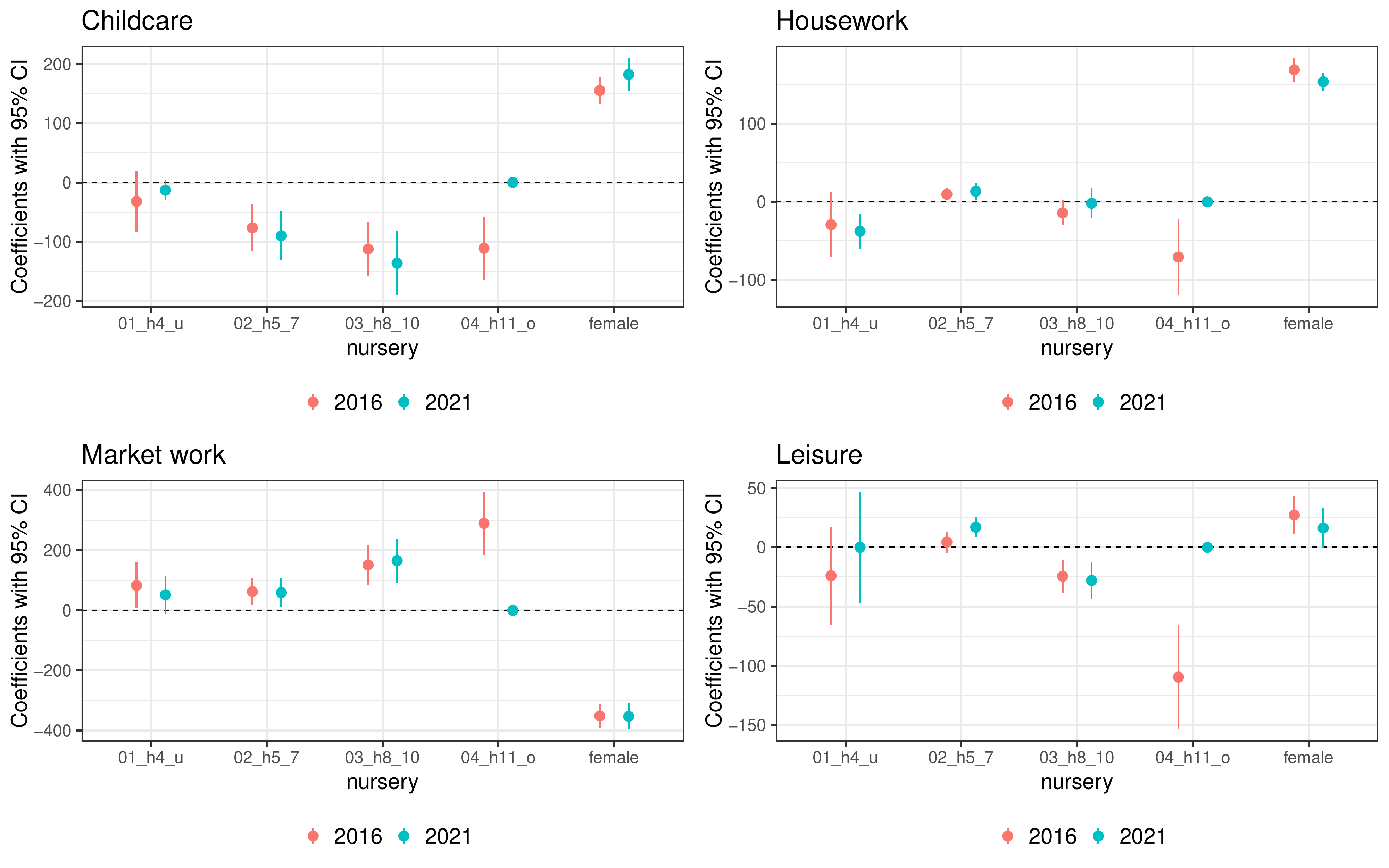}

\vspace{-0.2\baselineskip}

(c) Annual Earnings

\includegraphics[width=158mm, height=73mm]{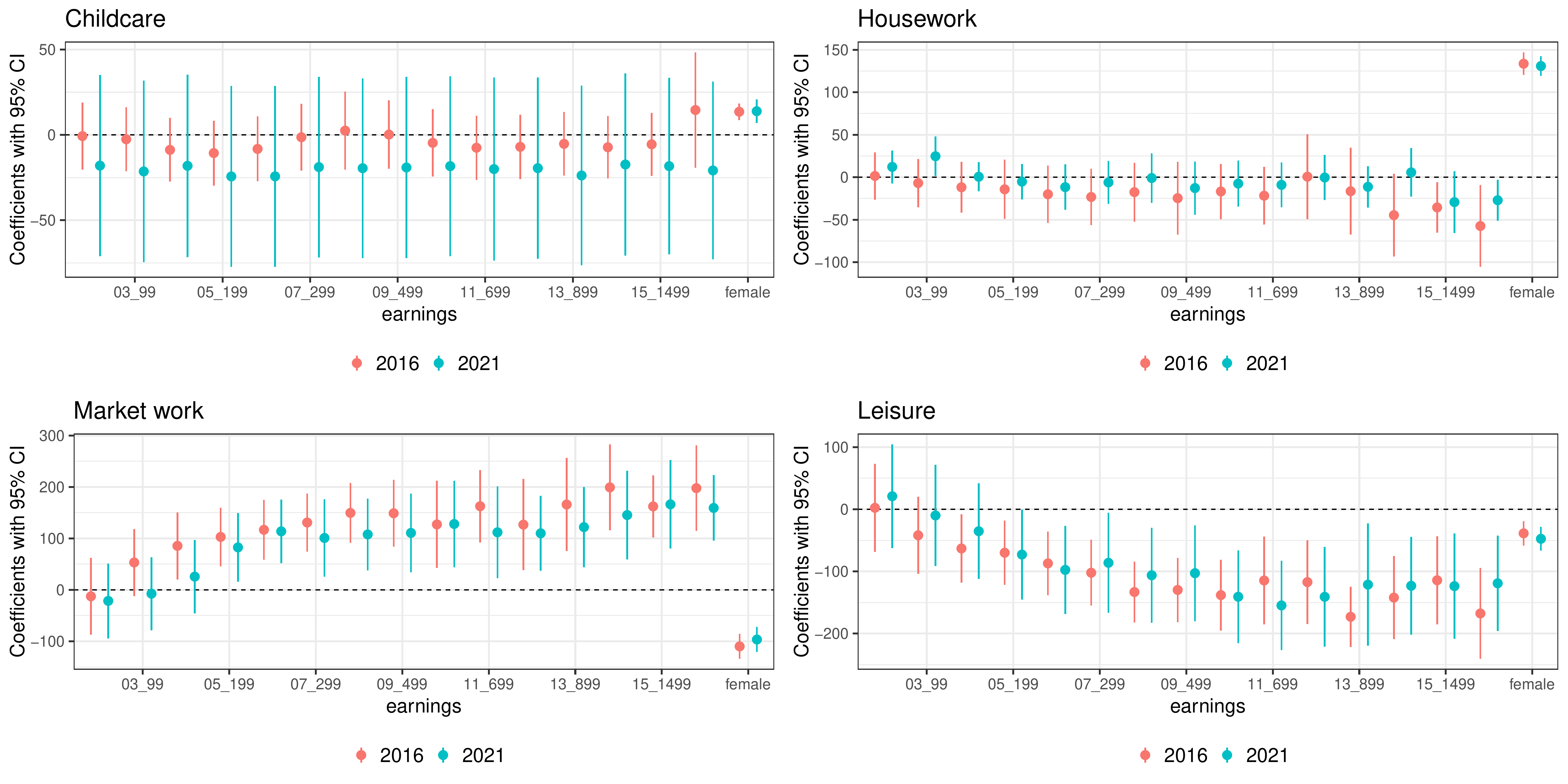}

\end{small}
\end{center}
\end{figure}

As additional supporting evidence, we will further investigate the relationship between parental time distribution and children's attendance at nursery schools or parental earnings through regression analysis. Figure \ref{fig:regression_time_allocation} displays the results of the regression analysis estimating parental time allocation, using data from the two most recent years, 2016 and 2021. In each graph, the markers and error bars represent OLS estimates and the corresponding 95\% confidence intervals, respectively.\footnote{We control all the regressions by the sample size corresponding to every group and its squared value as their regressors since the data are aggregated by groups.}

Panel (a) presents the regression results for four categories of childcare time allocation (measured in minutes per day) across eight groups based on the youngest child's age. These groups are: (1) age 0, (2) age 1--2, (3) age 3--5, (4) age 6--8, (5) age 9--11, (6) age 12--14, (7) age 15--17, and (8) age 18  and older. Age 0 serves as the baseline for the dummy variables. The control variables include the parent's gender and whether both parents are earners. The findings suggest that, as the youngest child's age increases, parents require fewer childcare hours, leading to more leisure time. However, no significant relationship is observed between the child's age and the parental market work or housework hours.

Panel (b) presents the regression outcomes concerning four categories of time allocation in relation to children's attendance at nursery schools. The time allocation categories are divided into five tiers: (1) no usage, (2) 4 hours or fewer, (3) 5--7 hours, (4) 8--10 hours, and (5) 11 hours or more. ``No usage'' is the baseline for the dummy variables, and the parent's gender and whether both parents are earners are control variables. The analysis shows that, when children spend more time at nursery school, parents spend less time on childcare and more time working in the market. However, increasing the amount of time that children spend at nursery school does not affect the amount of time that parents spend on housework or leisure.\footnote{Because there are few observed households using a nursery school for more than 11 hours in the 2021 data, their estimate is not shown in the graph.}

Panel (c) contains the regression results of parental time allocation in relation to individual annual income. Annual income is divided into 14 categories, such as 0, less than 0.5 million yen, 0.5--1.99 million yen, 2--2.49 million yen, 2.5--2.99 million yen, and seven subsequent categories ranging from 3 to 9.99 million yen, each separated by 1 million yen. There are also categories for 10--14.99 million yen and 15 million yen or higher. The control variables include the parent's gender. The outcomes show that an increase in parents' income leads to more market working hours and less leisure time. However, changes in annual income do not show a clear correlation with childcare or housework time, indicated by the inclusion of 0  within the error margins.

In summary, the empirical results for the most recent years, 2016 and 2021, highlight the significant impact of the age of the youngest child on parental time allocation. This emphasizes the necessity of constructing a model that effectively incorporates the relationship between a child's age and parental time allocation, thereby providing insights into household decisions throughout the life cycle.

\section{Model}
\label{sec:MODEL}

We use a life-cycle model incorporating the child's growth period from age 0 to age 18 into the utility function with a family structure, including childcare, which was developed by \citet{blundellChildrenTimeAllocation2018}. In this section, we describe first the unitary model, and then the model extends to a life cycle.\footnote{\citet{Almas2023} produced a survey paper that explains the utility function consisting of the members of a family.}

\subsection{Family Model with Childcare}


Let the utility function of a married couple with childcare be set up as follows:
\begin{align}
u(\cdot) =& \exp(\Phi_{c}) \ \frac{C^{1-1/\eta}}{1-1/\eta} - \frac{1}{1-\rho_{L}}\left[\exp(\Phi_{L1}) \ L_{1}^{1-1/\varphi_{L1}}+\exp(\Phi_{L2}) \ L_{2}^{1-1/\varphi_{L2}}\right]^{1-\rho_{L}} \notag\\
&-\frac{1}{1-\rho_{T}}\left[\exp \left(\Phi_{T1} \left(j_c \right) \right) \ T_{1}^{1-1/\varphi_{T1}}+\exp \left(\Phi_{T2} \left(j_c \right) \right) \ (T_{2}+ N )^{1-1/\varphi_{T2}}\right]^{1-\rho_{T}}
\end{align}
where $C$ is consumption, $L$ is leisure time, and $T$ is parenting time. The subscripts $1$ and  $2$  denote the husband's and the wife's time. $N$ in the last term on the right-hand side denotes their child's time spent at nursery school and increases the couple's utility, as does the mother's childcare time. $\Phi_c$,  $\Phi_L$, and $\Phi_T$ are the shift parameters for consumption, leisure, and parenting, respectively, indicating the strength of these preferences. Note that the shift parameter for parenting is assumed to depend on the child's age, $j_c$. 

The values of $\eta$,  $\psi_{L1}$, $\psi_{L2}$, $\psi_{T1}$, and $\psi_{T2}$ are the marginal rates of substitution for the preferences corresponding to consumption, leisure, and parenting of a husband and a wife. $\rho_L$  and  $\rho_T$ represent the shared relationship between the married couple. In cases in which these values are positive, they represent a complementary relationship between the married couple, whereas, in cases in which they are negative, they represent a substitutive relationship. In other words, when the values are positive, the married couple increases its utility when the husband and wife spend time together. Conversely, when they are negative, only one spends more time on childcare, which also increases the utility of the other. The estimation results of \citet{blundellChildrenTimeAllocation2018} show that $\rho_L$, indicating the share of leisure time, is positive, while  $\rho_T$, meaning the share of childcare time, is negative. Our calculation follows them.

\subsection{Life-Cycle Model}

The following Bellman equation is set up as a life-cycle model. The married couple decides on six matters: (1) consumption, (2) leisure time, (3) childcare time in the current period,  (4) assets in the next period,  (5) whether to have a baby at age 30 ($K=1$ or $0$), and  (6) whether to take parental leave ($PL=1$ or $0$).
\begin{align}
V\left(a, z_{1}, z_{2}, e_{1}, e_{2}, j\right) =& \max_{c, L, T, a^{\prime}, K, PL} u(\cdot) + s_{j} \beta \mathbb{E} \left[V \left(a^{\prime}, z_{1}^{\prime}, z_{2}^{\prime}, e_{1}^{\prime}, e_{2}^{\prime},\, j+1 \right) \mid z_{1},z_{2}\right] \notag\\
& + (1-s_{j}) \ \beta \ \mathbb{I}(j > Jr + 10) WG
\end{align}
where $u(\cdot)$ is the utility function of the unitary model described above. $\beta$ is the discount factor. $j$ is the age of the parents.
To simplify the model, parents are assumed to be the same age; $Jr$ is the retirement age; $s_j$ is the survival probability of the married couple, and it indicates that the couple will die at the same time. $z_1$ and $z_2$ are permanent idiosyncratic shocks in the married couple's labor productivity, respectively, whereas $e_1$ and $e_2$ are temporary idiosyncratic shocks to labor productivity.
These shocks represent future uncertainty. The magnitude of uncertainty is supposed to affect lifetime planning significantly. $\mathbb{I}(\cdot)$ is an indicator function, which returns 1  if the condition inside the parentheses is satisfied; otherwise, it returns 0. $WG$ is their warm-glow bequest for their children 10 years after retirement.

The budget constraint is set up with two equations before and after the retirement age $Jr$, respectively, as follows.
If $j<Jr$ and the couple chooses the use of a nursery school, then
\begin{align}
c+ f \times \mathbb{I}(K = 1,2 \le j_c \le 6) + a^{\prime} =& (1+r)a+w (1-\tau)\kappa_{j,1}z_{1,}e_{1}(H_{j,1}-H^{hw}_1) \notag{}\\ 
&+ w (1-\tau) \kappa_{j,2}z_{2}e_{2}(H_{j,2}-H^{hw}_2) + y \times \mathbb{I}(PL=1, j_c<2),
\end{align}
if {$j\ge Jr$} , then
\begin{align}
c+a^{\prime}=(1+r)a+pen
\end{align}

where $a$ is the married couple's assets, $r$ is the interest rate, $w$ is the wage rate, and $\kappa_{j,1}$ and $\kappa_{j,2}$ the labor productivity of husbands and wives at each age $j$. 
$H_{j,1}$ and $H_{j,2}$ are hours worked which can be divided into market work and housework: for $H^{hw}_i$, assuming that housework time is fixed across ages, we omit the subscript $j$ from $H^{hw}_i$ for $i=1,2$. 
$f$ denotes the cost of using a nursery school, and $y$ is the income benefit for taking PL. If parents do not use a nursery school, then $f$ is dropped from their budget. 
$pen$ is the pension of the household after retirement.

The sum of individual $i$'s leisure time at each age $j$: $L_{j,i}$, work time: $H_{j,i}$, and childcare time: $T_{j,i}$, is constant at 24 hours: $\overline{L}_j$. 
\begin{align}
L_{j,i}+H_{j,i}+T_{j,i}=\overline{L}_{j}, \ \ \ \mbox{for} \ \  i=1,2.
\end{align}
Next, two types of idiosyncratic shocks: $z_i$, $e_i$, which represent individual labor productivities of a husband: $i=1$ and a wife: $i=2$, are expressed as follows.
\begin{align}
\left[\begin{array}{c}
z_{1}^{\prime}\\
z_{2}^{\prime}
\end{array}\right]=\left[\begin{array}{cc}
\rho_{11} & \rho_{12}\\
\rho_{21} & \rho_{22}
\end{array}\right]\left[\begin{array}{c}
z_{1}\\
z_{2}
\end{array}\right]+\left[\begin{array}{c}
\epsilon_{1}\\
\epsilon_{2}
\end{array}\right],\,\,\,\,\,\,\left[\begin{array}{c}
\epsilon_{1}\\
\epsilon_{2}
\end{array}\right]\sim N(0,\Sigma^{\epsilon}),
\end{align}
\begin{align}
\left[\begin{array}{c}
e_{1}\\
e_{2}
\end{array}\right]\sim N(0,\Sigma^{e}).
\end{align}
The former persistent labor productivity, $z_{i}$, has continuity from past productivity, which economically signifies the accumulation of human capital known as a career. Therefore, the correlation of labor productivity between spouses is set to be extremely weak, that is, $\rho_{12}=\rho_{21}=0$. Conversely, the latter temporary productivity shock, $e_{i}$, pertains to the environmental factors affecting the couple at each point in time, and it is assumed that the productivity of the two individuals is highly positively correlated, specifically $\sigma_{12}=\sigma_{21} > 0$.\footnote{The environmental factors refer to exogenous shocks related to living arrangements, such as housing, food, and clothing, that arise from the cohabitation of a married couple. These shocks affect both spouses mutually but are temporary in duration.}

\section{Methods and Parameter Setting }
\label{sec:PARAMETER_SET}
\subsection{Numerical and Estimation Methods }
This section briefly outlines the numerical computation and estimation methods employed to obtain the equilibrium solutions of the model discussed in this paper. As will be detailed below, these two methods have been adopted from the latest existing research. We consider that these two methods are satisfactory in terms of computational accuracy and robustness of estimations.

The Value Function Iteration (VFI) Toolkit by \citet{kirkbyToolkitValueFunction2017} is set for the numerical computation in our study. We use the tool by \citet{kirkbyLifeCycleModels2022} to calculate our life-cycle model. The advantage of this tool is that it implements fast parallel computing by using a Graphical Processor Unit (GPU) in addition to a Central Processor Unit (CPU). The value function that we are computing has six dimensions and will suffer from the so-called ``curse of dimensionality.'' GPUs are a powerful tool that can overcome this drawback.

For calculating the value function of the Bellman equation, we have the five control variables determined by the agents. The number of grids for leisure time and parenting time is 11 each, and two grids each are assigned to childbirth and PL. For the state variables, we have 51 grids for assets, three grids for the persistent idiosyncratic shocks to individuals in the couple, and three grids for two temporary idiosyncratic shocks. The married couple is assumed to start at age 20 and survive to a maximum of age 100, so the number of grids is 81. Thus, the total number of grids in the value function is $51 \times 3^2  \times 3^2 \times 81 = 334,611$.

We employ the GMM, as used by \cite{Greenwood2016}, to estimate the parameters of the heterogeneous agent model from \textit{aggregated data}. Following the approach of \citet{blundellChildrenTimeAllocation2018}, we estimate 14 parameters, setting the preference shifter for consumption by childless couples, $\Phi_C$, as given, to serve as a baseline for normalizing other parameters. These parameters are described in Table \ref{tab:estimate_parameters}. For the GMM moments, we use three aggregate data from the STULA, as discussed in Section \ref{sec:DATA}, specifically the daily average time spent on work, leisure, and childcare over a 45-year span for wives aged 20 to 64. We use the deviations between the model-derived time allocations and aggregated work, leisure, and childcare data, as the three first moments of the GMM. The weights for these three GMM moments are the reciprocals of the daily average times for each activity, specifically, 1/8 for labor, 1/12 for leisure, and 1/4 for childcare. 

We assume that the female agents in our model consist of four types based on their educational attainment and whether they use a nursery school for their children or not.\footnote{In our study, we classified female agents into these four exogenously assigned types for the following reasons. Firstly, regarding educational background, we assumed that agents had already chosen their level of education—either college or high school—before entering the labor market. Although recent studies, such as those by \cite{Greenwood2016}, determine educational attainment based on innate ability levels and future income expectations, incorporating endogenous educational decision-making in our model would complicate it further, which we opted to avoid. Secondly, daycare use is largely determined by local government approval rather than parental choice in Japan.}\hspace{0.125em}\footnote{In Japan, the large number of children on waiting lists who cannot attend nurseries has become a serious issue. Currently, it is estimated that half of the children are enrolled in nurseries, while the remaining children likely include a significant number of those on waiting lists. These children are typically cared for by their grandmothers, through extended parental leave, or by their mothers leaving their careers. This study does not take into account the children on waiting lists.}
Each of the four types is set to make up 25\% of the total, as shown in Panel (a) of Table \ref{tab:set_parameters}, the values of which are based on the actual situation in Japan.\footnote{According to the ``Basic School Statistics'' by the Ministry of Education, Culture, Sports, Science, and Technology in 2023, the university enrollment rate was approximately 54\% for men and 51\% for women. Furthermore, the nursery utilization rate was derived from the 2023 ``Summary of Nursery School and Related Conditions.'' According to this report, the overall utilization rate for children of all ages was 52.4\% as of April 2023 and 50.9\% as of April 2022. Therefore, the values set in Panel (a) of Table \ref{tab:set_parameters} appear to reflect the current situation in Japan.}
Additionally, our model assumes that wives with either a high school or college education engage in \textit{assortative mating} with husbands of the same educational level.\footnote{Empirical analysis of assortative mating is considered by \cite{Greenwood2016} in terms of a unitary model in the US.}
Following the life-cycle model by \cite{Borella2018}, we construct aggregated data composited from the four types.

\subsection{Parameter Setting}

\begin{table}[tb]
\caption{Parameter Setting}
\label{tab:set_parameters}
\begin{center}
\begin{small}

(a) Ratio of Heterogenous Wives

\begin{tabular}{l|c} 
\hline 
Types of Wives & Ratio \\ \hline 
College Graduate and Using Nursery & 0.25 \\  
College Graduate and Not Using Nursery & 0.25 \\ 
High School Graduate and Using Nursery & 0.25 \\ 
High School Graduate and Not Using Nursery & 0.25 \\   \hline 
Total & 1.00 \\   \hline 
\end{tabular}

\bigskip{}

(b) Calibration Parameters

\begin{tabular}{c|c|l} 
\hline 
Parameters & Values & Explanation \\ \hline
$\beta$ & 0.96 & Discount rate  per year\\
$r_t$ & 0.05 & Interest rate  per year \\ 
$J$ & 20 & Age of parental entry into the market \\ 
$Jr$ & 65 & Retirement age \\ 
$H^{hw}_i$ & 0.125 & Housework: 3 hours per day \\
$N$ &  0.167 & Nursery time spent: 4 hours per day \\
$y$ & $0.5w(1-\tau)\kappa_{j,2}(H_{j,2}-H^{hw}_2)$ & PL's income compensation \\ 
$f$ & $0.25w(1-\tau)\kappa_{j,2}(H_{j,2}-H^{hw}_2)$ & Cost of nursery per year \\ 
$pen$ & $0.3w(1-\tau)\biggl( \kappa_{Jr,1}(H_{Jr,1}-H^{hw}_1)+\kappa_{Jr,2}(H_{Jr,2}-H^{hw}_2) \biggr)$ & Pension per household \\ 
$w$ & $1$ & Real wage per year \\
\hline 
\end{tabular}

\bigskip{}
(c)	Idiosyncratic Shocks

\begin{tabular}{cc|cc|cc} 
\hline 
\multicolumn{4}{c|}{AR(1) shocks}   & \multicolumn{2}{c}{Temporary shocks}  \\
Parameters & Values & Parameters & Values & Parameters & Values \\
\hline
$\rho_{11}$ & 0.9 & $\sigma_{11}^{\epsilon}$ & 0.0303 & $\sigma_{11}^{e}$ & 0.1 \\
$\rho_{12}$, $\rho_{21}$ & 0 & $\sigma_{12}^{\epsilon}$, $\sigma_{21}^{\epsilon}$ & 0.0027 & $\sigma_{12}^{e}$, $\sigma_{21}^{e}$ & 0.05 \\
$\rho_{22}$ & 0.7 & $\sigma_{22}^{\epsilon}$ & 0.0382 & $\sigma_{22}^{e}$ & 0.1 \\ 
\hline 
\end{tabular}

\end{small}

\vspace{0.5\baselineskip}
    \begin{footnotesize}
    \end{footnotesize}
\end{center} 
\end{table}

As shown in Panel (b) of Table \ref{tab:set_parameters}, we define the configuration for the fixed parameters. The time discount rate, denoted as $\beta$, is set at 0.96, while the interest rate is established at 5\%. The retirement age, $Jr$, is determined as 65. The rate of income benefit provided during PL (income replacement rate) is designated as 50\%, and the nursery school fee is equivalent to one-quarter of the mother's income. Pension payments are stipulated to amount to 30\% of the working-age family's income. As shown in Figure \ref{fig:data_time_allocation}, mothers' housework hours are approximately 3--4 hours regardless of their age. Therefore, in our model, we set it to three hours per day. Next, the settings for idiosyncratic shocks are detailed in Panel (c) of Table \ref{tab:set_parameters}. In Japan, the career effect on women's labor income based on years of continuous service is smaller than that for men, with a value of 0.7 for $\rho_{22}$ compared with 0.9 for $\rho_{11}$ for men.

\subsection{Wages and  Survival Rates}
\label{subseq:data_demographics_wage}

We describe the fixed parameters for labor productivity and survival rates, which are derived from wage and demographic data. The demographic information is grounded in age- and gender-specific survival rates, while labor productivity is derived from gender- and education-based hourly wage data.

\begin{figure}
\caption{Data: Labor Productivity and Survival Rates}
\label{fig:data_LP_SR}

\begin{center}

\noindent\begin{minipage}[t]{1\columnwidth}%
\includegraphics[width=8cm,height=6cm]{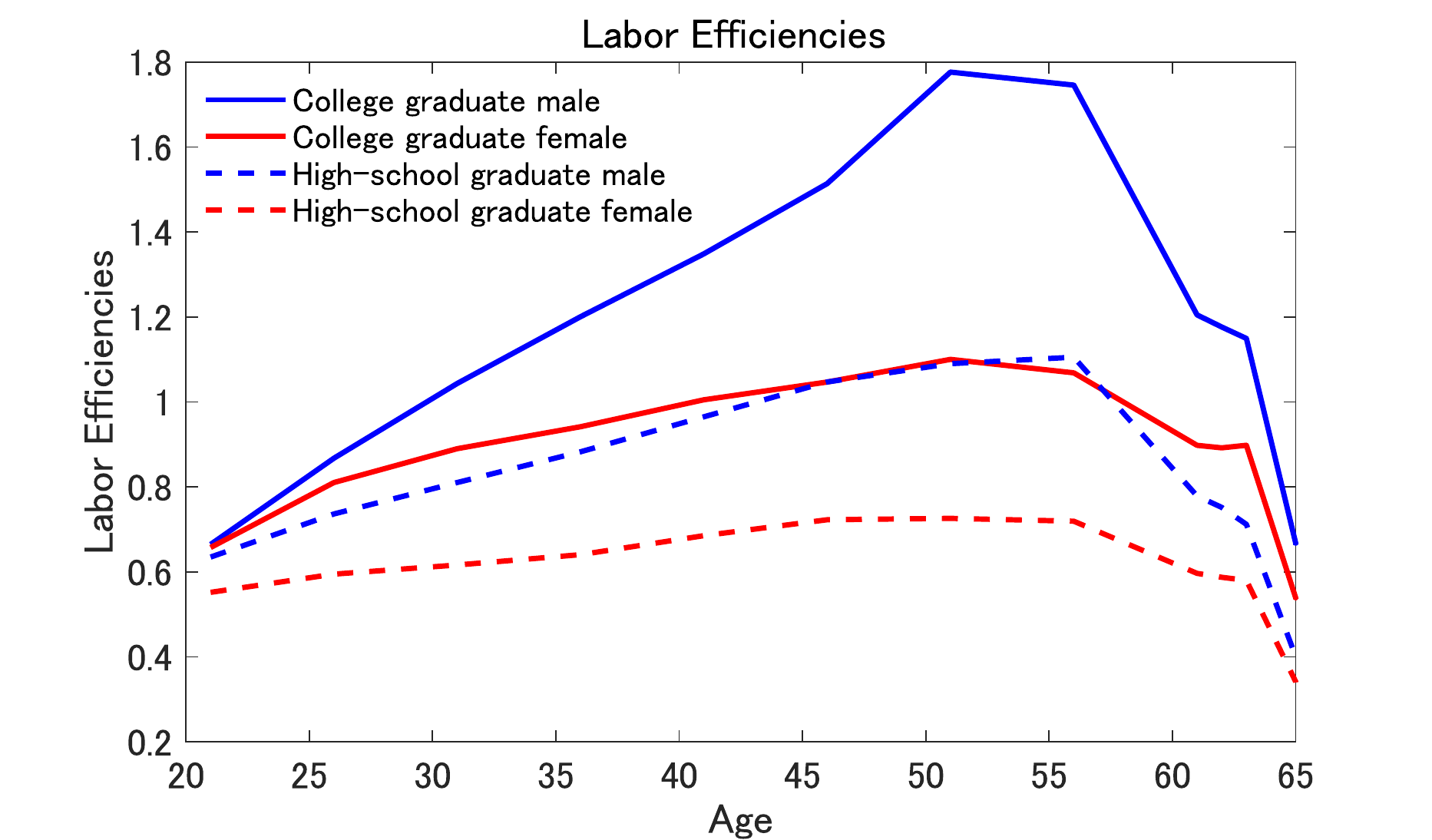}\includegraphics[width=8cm,height=6cm]{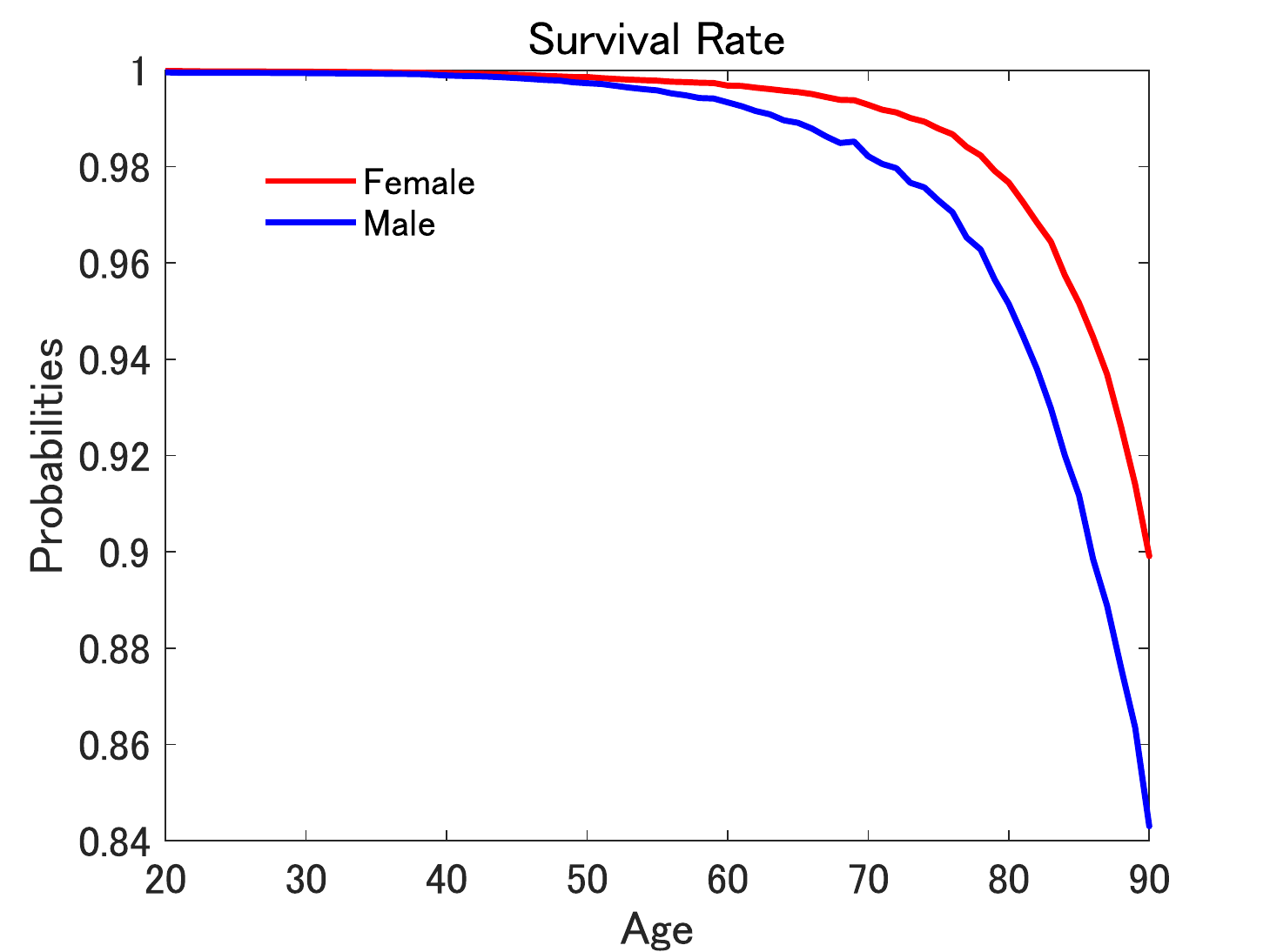}
\end{minipage}

\end{center}
\end{figure}

First, we describe the categories of agents by gender and educational attainment and the corresponding labor productivity for each category. These distinctions are established using the Basic Survey on Wage Structure Statistics (2019) provided by the Ministry of Health, Labour, and Welfare. To account for gender and educational diversity among the agents, we distribute the population across four categories in accordance with the composition ratios detailed in the dataset. By employing these data, we calculate the hourly wages pertinent to each agent category and subsequently gauge the labor productivity using the method introduced by \citet{Hansen1993} and employed by \citet{BIJ2006}, \citet{Yamada2011}, and others. Panel (a) of Figure \ref{fig:data_LP_SR} shows the age-related distribution of labor productivity between the ages of 20 and 65, classified by gender and educational attainment. The solid blue and red lines represent the labor productivity of men and women with more than a college education, while the dashed blue and red lines denote the labor productivity of males and females with an education level below that of a college degree.

Next, we consider agents distinguished by age- and gender-dependent survival rates, derived from the population distribution and mortality rate projections provided by the National Institute of Population and Social Security Research (IPSS; 2017 estimates). This dataset allows us to account for life cycle variations in discounts by determining gender-specific aggregated survival rates and integrating them with the discount rate within the Bellman equation. Panel (b) of Figure \ref{fig:data_LP_SR} shows that women exhibit consistently higher survival rates across the life cycle and that this gender disparity becomes more pronounced, particularly beyond the age of 60.

\subsection{Estimation of Parameters}
The parameters estimated in our model correspond to those identified by \citet{blundellChildrenTimeAllocation2018}. Table \ref{tab:estimate_parameters} presents a comparison between our estimates and those of \citet{blundellChildrenTimeAllocation2018}. There are two important considerations to note here. First, the data that we use for the estimation are aggregated, whereas \citet{blundellChildrenTimeAllocation2018} utilized individual-level data. Second, our model is a life-cycle model in which both parents and children are assigned specific ages, in contrast to their unitary model, which abstracts away the ages of agents. Panel (a) of Table \ref{tab:estimate_parameters} documents the estimated values for each marginal rate of substitution (MRS) parameter. A positive value of $\rho_L$ indicates that married couples share leisure time, which contributes to their overall utility. Conversely, a negative value of $\rho_T$ suggests that couples tend to divide the work of childcare rather than performing it together.

\begin{table}[tb]
\caption{Estimation of Parameters}
\label{tab:estimate_parameters}
\begin{center}

\begin{footnotesize}
(a) MRS Parameters

\begin{tabular}{c|c|c|c|c|c}
\hline 
\cline{2-6} \cline{3-6} \cline{4-6} \cline{5-6} \cline{6-6} 
 \multicolumn{3}{c|}{Leisure and Consumption} &   \multicolumn{3}{c}{Parental Time}\tabularnewline
 & Our Estimation & \citet{blundellChildrenTimeAllocation2018} &  & Our Estimation & \citet{blundellChildrenTimeAllocation2018}\tabularnewline
 \hline
$\psi_{L1}$ & 0.1986 & 0.211 & $\psi_{T1}$ & 0.2067 & 0.115\tabularnewline
 
$\psi_{L2}$ & 0.1518 & 0.162 & $\psi_{T2}$ & 0.5440 & 0.503\tabularnewline
 
$\rho_{L}$ & 0.4745 & 0.535 & $\rho_{T}$ & -0.1878 & -0.197\tabularnewline
$\eta$ & 0.4693 & 0.903 &  &  & \tabularnewline
\hline 
\end{tabular}

\bigskip{}

(b) Preference Shifters

\begin{tabular}{c|c|c|c|c}
\hline 
\cline{2-5} \cline{3-5} \cline{4-5} \cline{5-5} 
 & \multicolumn{2}{c|}{With Children} & \multicolumn{2}{c}{Without Children}\tabularnewline
 & Our Estimation & \citet{blundellChildrenTimeAllocation2018} & Our Estimation & \citet{blundellChildrenTimeAllocation2018}\tabularnewline
\hline 
$\Phi_{L1} (j=Age 20)$  & $-$8.7970 & $-$8.925 & $-$6.4374 & $-$7.680\tabularnewline
 
$\Phi_{L2} (j=ge 20)$ & $-$8.8182 & $-$9.397 & $-$8.8798 & $-$8.816\tabularnewline
 
$\Phi_{T1}(j=Age 30)$ & $-$20.9558 & $-$23.993 & na & na\tabularnewline
 
$\Phi{}_{T2}(j=Age 30)$ & $-$3.4116 & $-$3.957 & na & na\tabularnewline
 
$\Phi_{C}$ & $-$0.9785 & 0.132 & $-$1.0 (fixed) & 0 (fixed) \tabularnewline
\hline 
\end{tabular}

\end{footnotesize}
\end{center}

\vspace{0.5\baselineskip}
    \begin{footnotesize}
Note: (1) The preference for leisure time is set larger for husbands than for wives in married couples over age, i.e., $\Phi_{L, j, 1} > \Phi_{L, j, 2}$. (2) Wives have a greater preference for childcare time than leisure time, i.e., $\Phi_{T2} > \Phi_{L2}$. (3) Wives also have a more extensive selection for childcare than husbands, i.e., $\Phi_{T2} > \Phi_{T1}$. (4) Childcare preferences are assumed to be determined by the child’s age, with a maximum at the child’s age 0 (or parental age 30) and decreasing linearly until the child reaches age 18, as shown in Figure \ref{fig:time_varying_preference}.
\end{footnotesize}

\end{table}

In Panel (b), it can also be seen that the preference shift parameter shows a value close to the estimates of \citet{blundellChildrenTimeAllocation2018}. The value of the preference parameter $\Phi_{L,j,1}$ for the husband's leisure is greater than that for the wife $\Phi_{L,j,2}$, which means that the former prefers leisure more than the latter.
Similarly, wives exhibit a stronger inclination to allocate time to childcare than leisure, signifying that $\Phi_{T,j,2} > \Phi_{L,j,2}$. Furthermore, wives demonstrate a more extensive preference for childcare than husbands, indicating that $\Phi_{T,j,2} > \Phi_{T,j,1}$. 
In contrast to \citet{blundellChildrenTimeAllocation2018}, we employ a life-cycle model, allowing parameters $\Phi_{L,j,i}$, $\Phi_{T,j,i}$, such as leisure and childcare time, to vary depending on the age of the children and parents.

\begin{figure}[tb]
\caption{Age-Dependent Variations of the Preference Shifter}
\label{fig:time_varying_preference}
\begin{center}

\noindent\begin{minipage}[t]{1\columnwidth}%

\includegraphics[width=8cm,height=7cm]{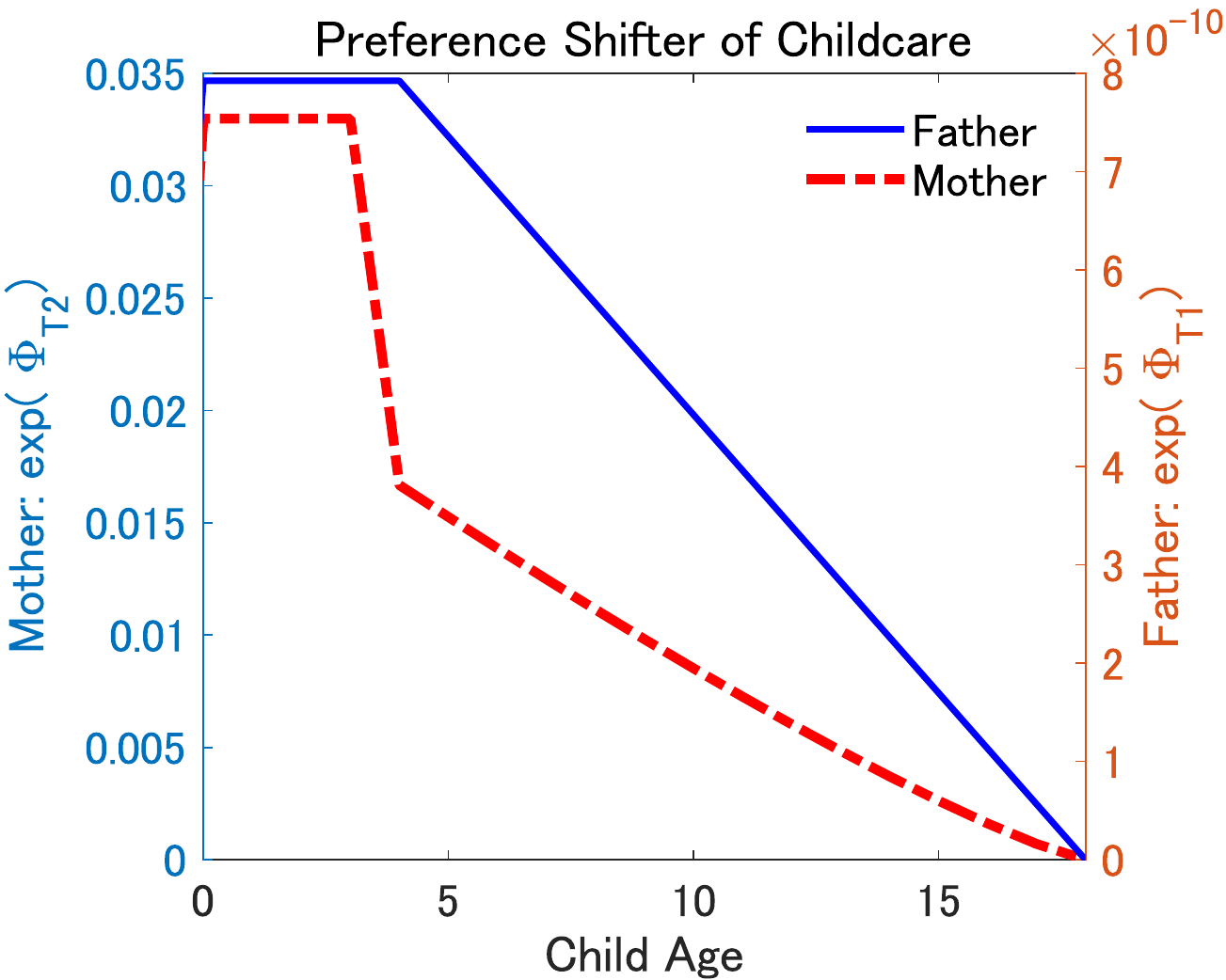} \includegraphics[width=8cm,height=7cm]{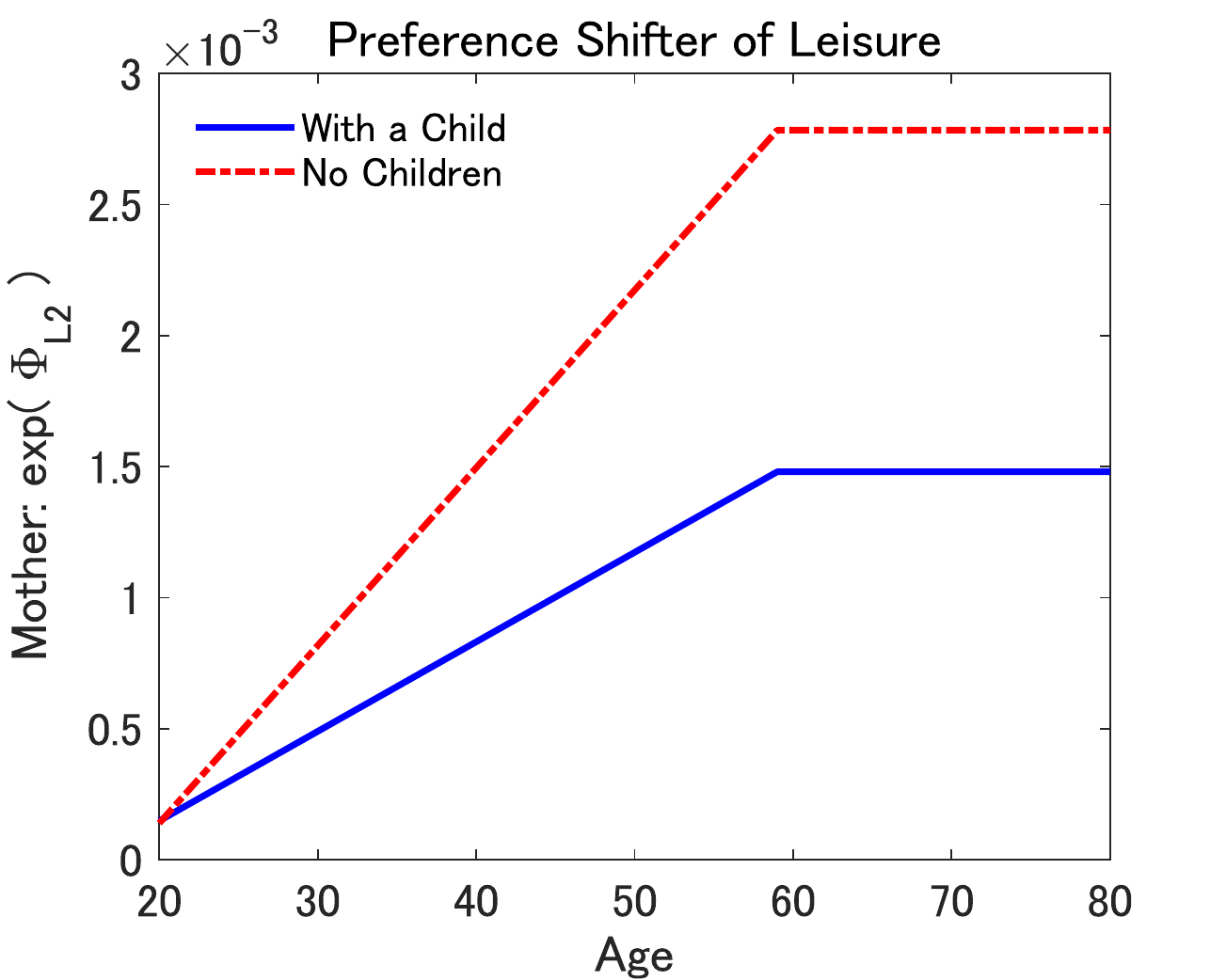}%
\end{minipage}

\end{center}
\end{figure}

As shown in the left graph of Figure \ref{fig:time_varying_preference}, the parental preference for childcare peaks when the child is between 0 and 3 years old and then gradually decreases until the child reaches 18 years old. Additionally, as shown in the right graph of Figure \ref{fig:time_varying_preference}, the preference for leisure increases uniformly until retirement at age 65, with the magnitude varying depending on whether the parents have children.

\section{Estimation Results}
This section discusses the results of our life-cycle model calculations and the validity of the model, using the estimated parameters in Table \ref{tab:estimate_parameters}. First, we describe the results of our calculations of parental decisions by asset level and age. We then show the evolution of time allocation decisions over the life cycle. Finally, we discuss the deviations between our results and the data, focusing on time allocation and child penalties.

\label{sec:NUMEARICAL}
\subsection{Calculations of Policy Functions}

\begin{figure}
\caption{Heat Map of Decisions of Heterogenous Agents by Ages and Assets}
\label{fig:policy_func}

\begin{center}

(a) Working Time of Women

\includegraphics[width=12.5cm]{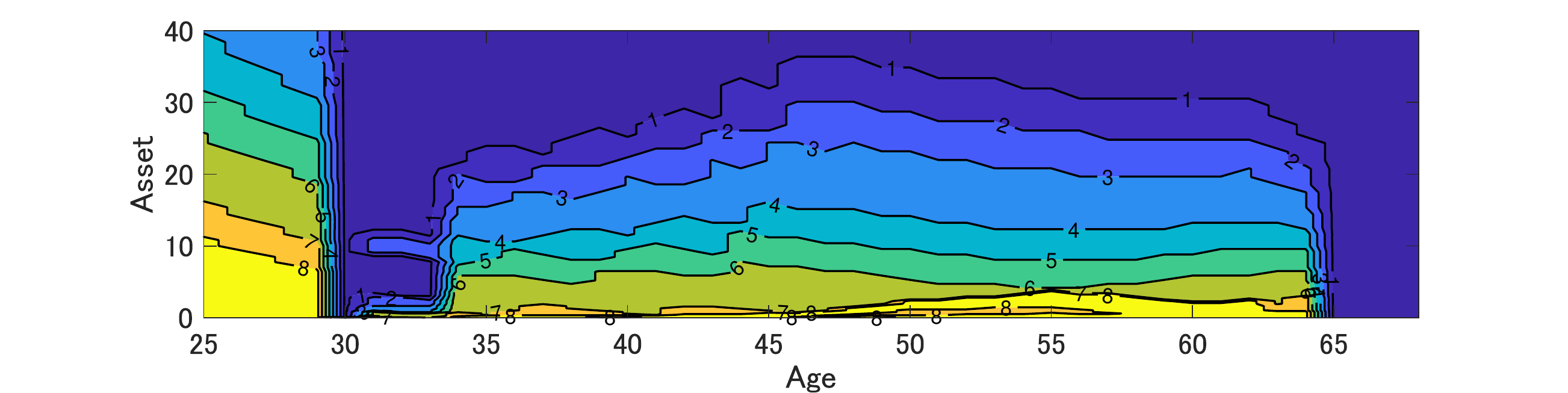}

\bigskip{}

(b) Childcare Time of Mothers Using a Nursery

\includegraphics[width=12.5cm]{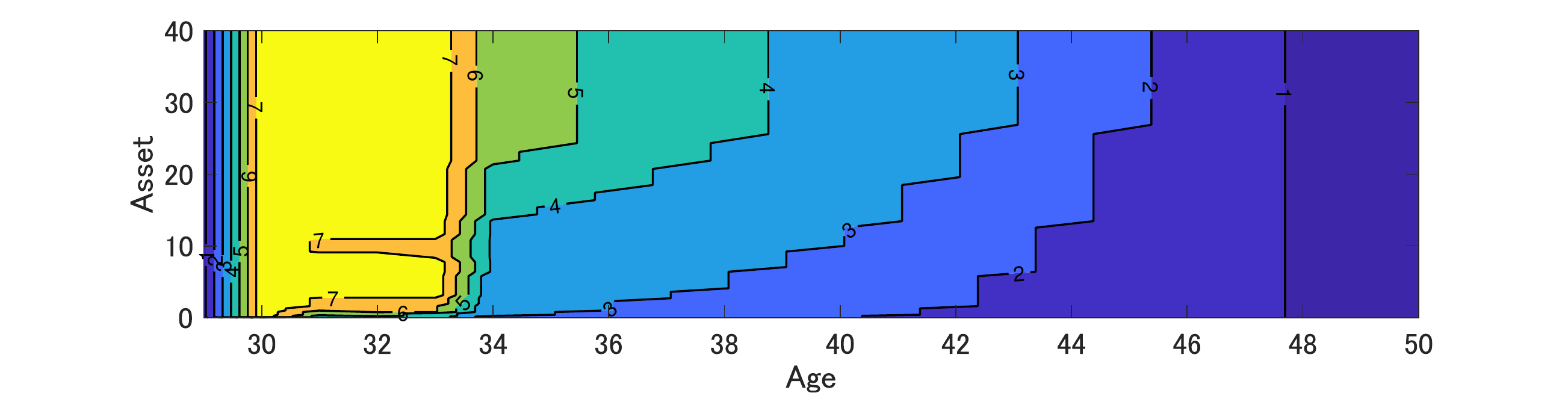}
\bigskip{}

(c) Decision on Parental Leave of Mothers

\includegraphics[width=12.5cm]{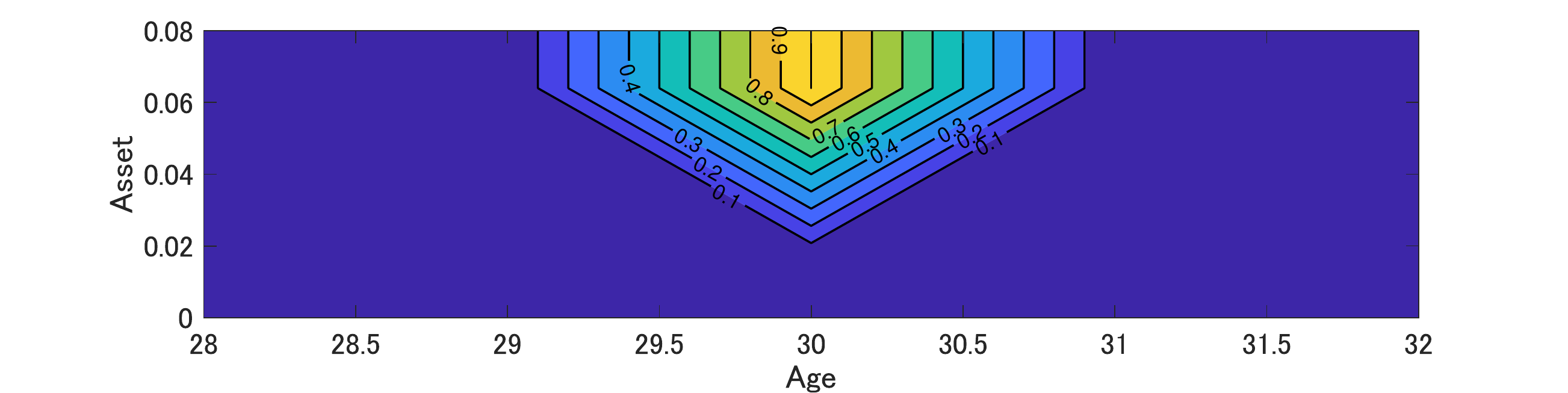}

\bigskip{}

(d) Distribution of Assets of Families

\includegraphics[width=12.5cm]{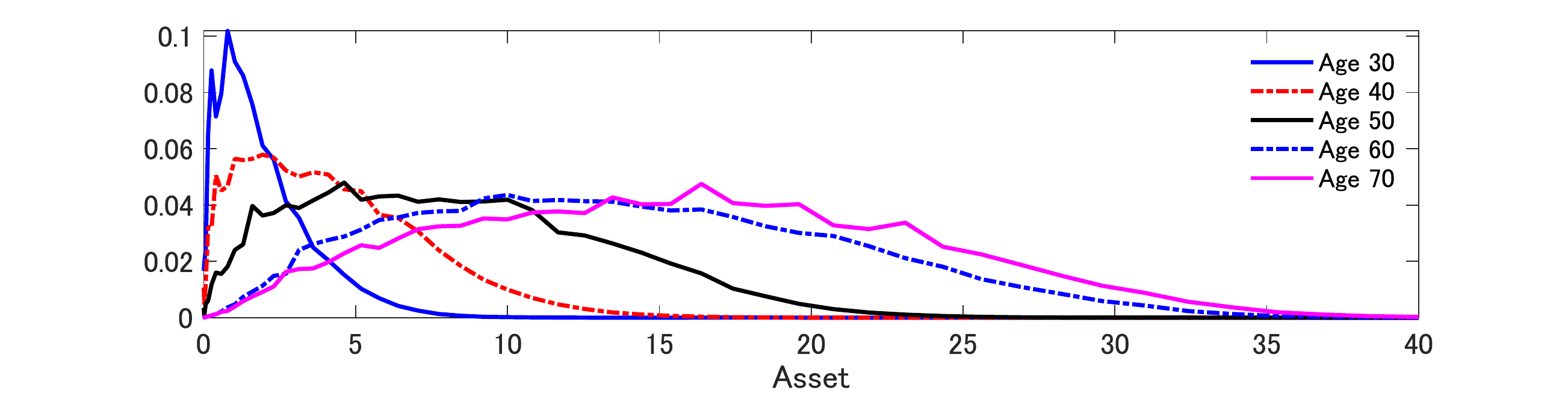}

\end{center}
\end{figure}

Figure \ref{fig:policy_func} 
illustrates the policy function derived from numerical computation with the estimates of Table \ref{tab:estimate_parameters}. The horizontal axis represents the parental age, while the vertical axis indicates the household's asset level. Panel (a) provides insights into the determination of women's working hours. As a family's assets increase, the wife's working hours decrease correspondingly. In contrast, Panel (b) shows women's choices regarding parenting time. With higher assets, more time is allocated to childcare responsibilities. Panel (c) describes the decision-making process for PL. Notably, PL is not taken when the household's assets fall below 0.05. Panel (d) depicts the distribution of assets across different age groups. We observe a limited accumulation of assets within the 30 age group. However, beyond the 50 age group (the black solid line), the asset distribution widens significantly for different households of the same age.

\subsection{Time Allocation throughout the Life cycle}

\begin{figure}
\caption{Average of Decisions of Heterogenous Agents by Age}
\label{fig:result_average}

\begin{center}

(a) College Graduate Parents Using a Nursery School

\includegraphics[width=16cm]{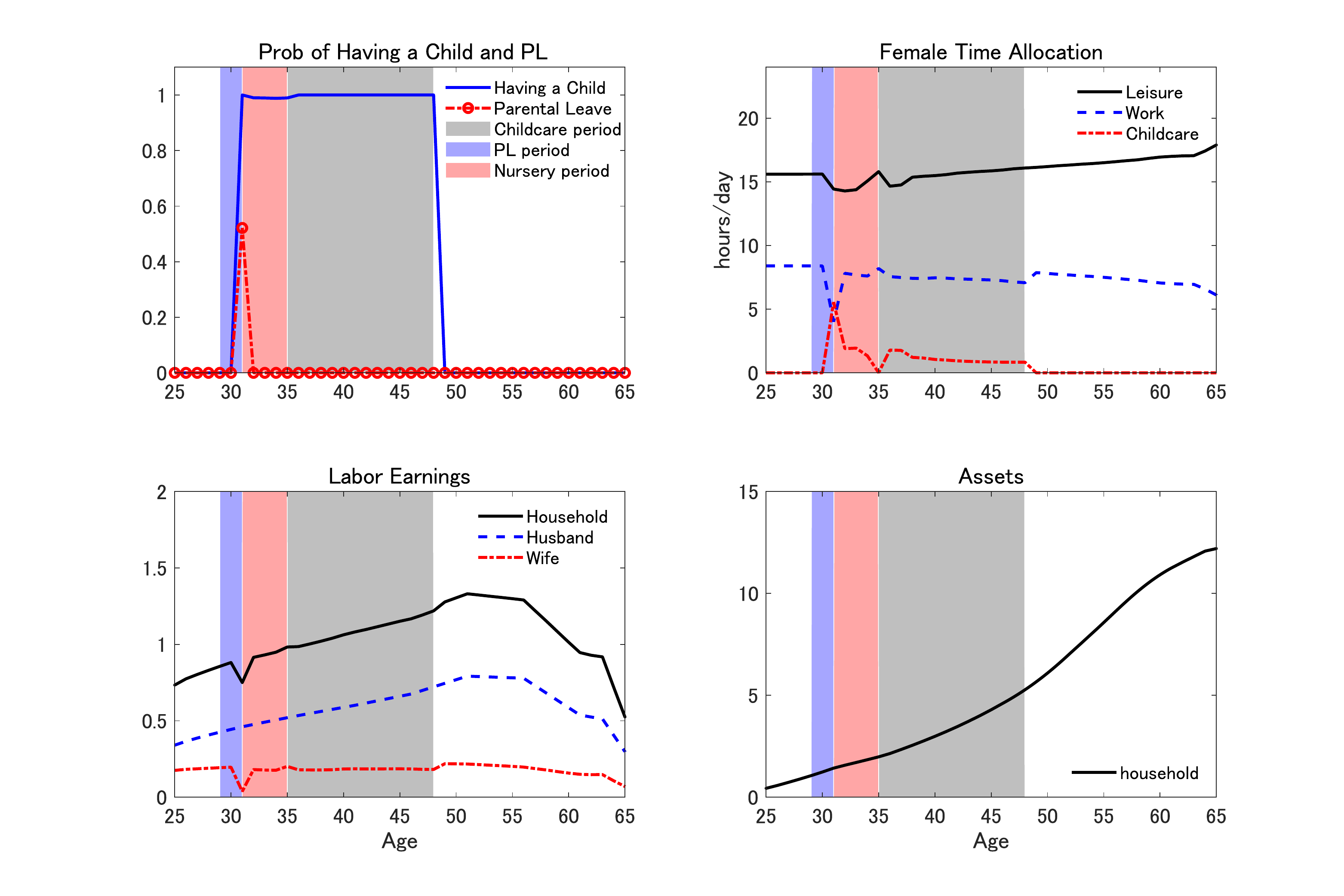}

\bigskip{}

(b) High-School Graduate Parents Not Using a Nursery School

\includegraphics[width=16cm]{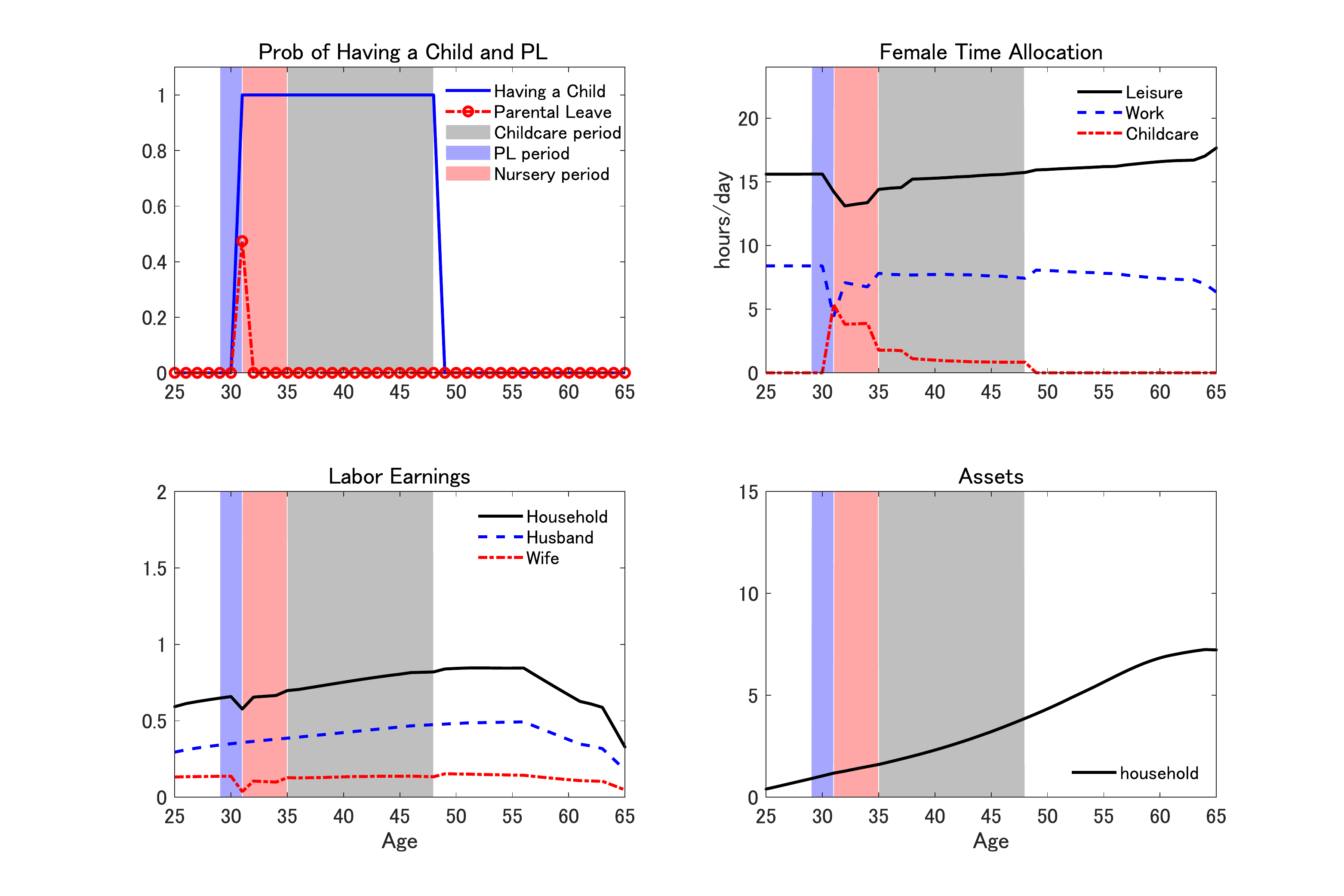}

\end{center}
\end{figure}

In Panel (a) of Figure \ref{fig:result_average}, the calculations focus on a scenario involving a married couple categorized as ``college graduates'' who use a nursery school. This calculation sets the income RR for PL at 50\%. The values shown in all the graphs represent averages derived from various individuals within each parental age group. The graph in the top left corner displays the probability of taking PL. The top right graph visually displays how the mother divides her time among leisure, work, and childcare. The bottom left graph illustrates the combined labor income for the husband, wife, and total of the married couple. Lastly, the bottom right graph depicts the total assets of the married couple.

In the upper left plot of Panel (a), our model reveals that diverse individuals have an almost 100\% higher probability of choosing to have a child at the age of 30. Furthermore, they exhibit approximately a 50\% increased probability of taking PL. This underscores that, for almost all individuals, the utility function value resembling that proposed by \citet{blundellChildrenTimeAllocation2018} is greater when choosing to have children than when not choosing to have them. In the top right graph, the red dashed line represents the average time allocated to childcare by different individuals. This line shows a significant decrease in maternal childcare time as PL ends and the child starts attending nursery school aged 2 (the light red shade). Subsequently, the childcare time increases from the child's age of 5, corresponding to the end of nursery school usage (the light gray shade). Conversely, the blue dashed line illustrates the maternal working hours. Notably, there is a significant reduction in working hours during the first year after childbirth, followed by a rise again shortly thereafter. This suggests that mothers spend less time on childcare due to the use of the day nursery school and use this time for work instead.

The labor earnings depicted in the lower left graphs are calculated based on gender- and educational attainment-specific labor productivity, as detailed in Section \ref{sec:DATA}. Following childbirth, women work for approximately 6 hours per day, fewer than the 8 hours per day for men, as shown in Figure 1. The earnings of women and female labor earnings, represented by the red dashed line, exhibit a relatively steady trajectory with increasing age. In contrast, male labor earnings consistently rise until the age of 55, regardless of whether men have children. These computational findings align with the empirical evidence of the ``child penalty'' phenomenon, which has recently gained attention among labor economists \citep[e.g][]{Kleven2019a, Kleven2023}.

In the lower right graph of Figure 6, the accumulation of household assets is visualized. Assets reach their peak at age 65, with stability primarily attributed to pension disbursements and interest income from assets.

The graph in Panel (b) illustrates the case of a couple with a high school education who do not use a nursery. We will discuss later how differences in education level and nursery school use affect time allocation. For now, we will briefly explain the contents of the graph. The two graphs on the bottom left and right, respectively, depict changes in earnings and assets over time. The income and assets of individuals with a high school education are approximately two-thirds of those with a college education. Conversely, during the mother's childcare period (the light red shade), both the extended childcare time due to the non-use of a nursery and the decrease in earnings associated with having a high school education significantly reduce the mother's leisure time (the black solid line) in the upper right graph.

\subsection{Model Validation for Female Time Allocation }
\begin{figure}
\caption{Model Validation for Female Time Allocation}
\label{fig:result_model_validation}

\begin{center}

~~~~~~~~~~(a)  College Graduate Parents Using a Nursery

\includegraphics[width=13.5cm]{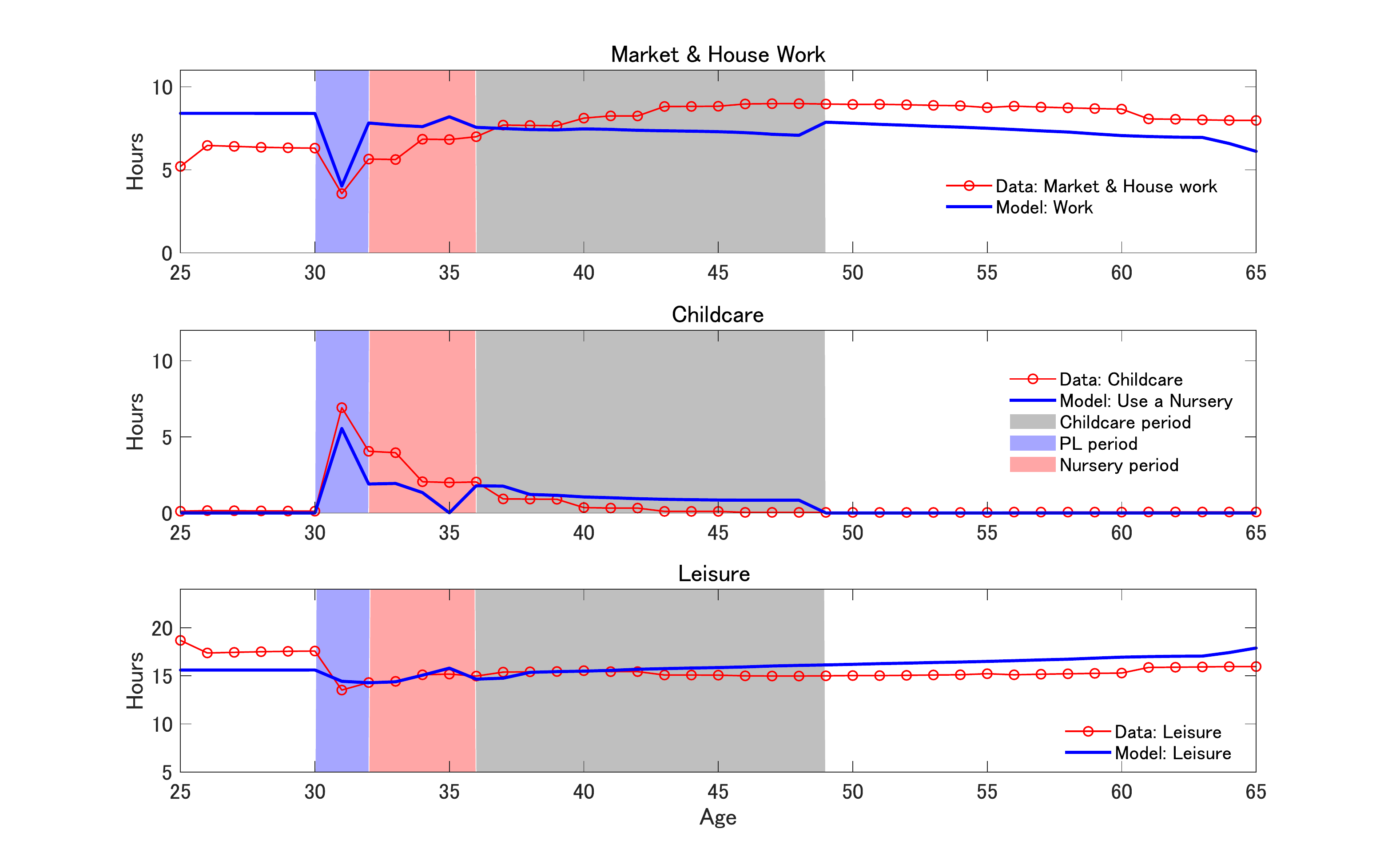}

\bigskip{}

~~~~~(b)  High-School Graduate Parents Not Using a Nursery

\includegraphics[width=13.5cm]{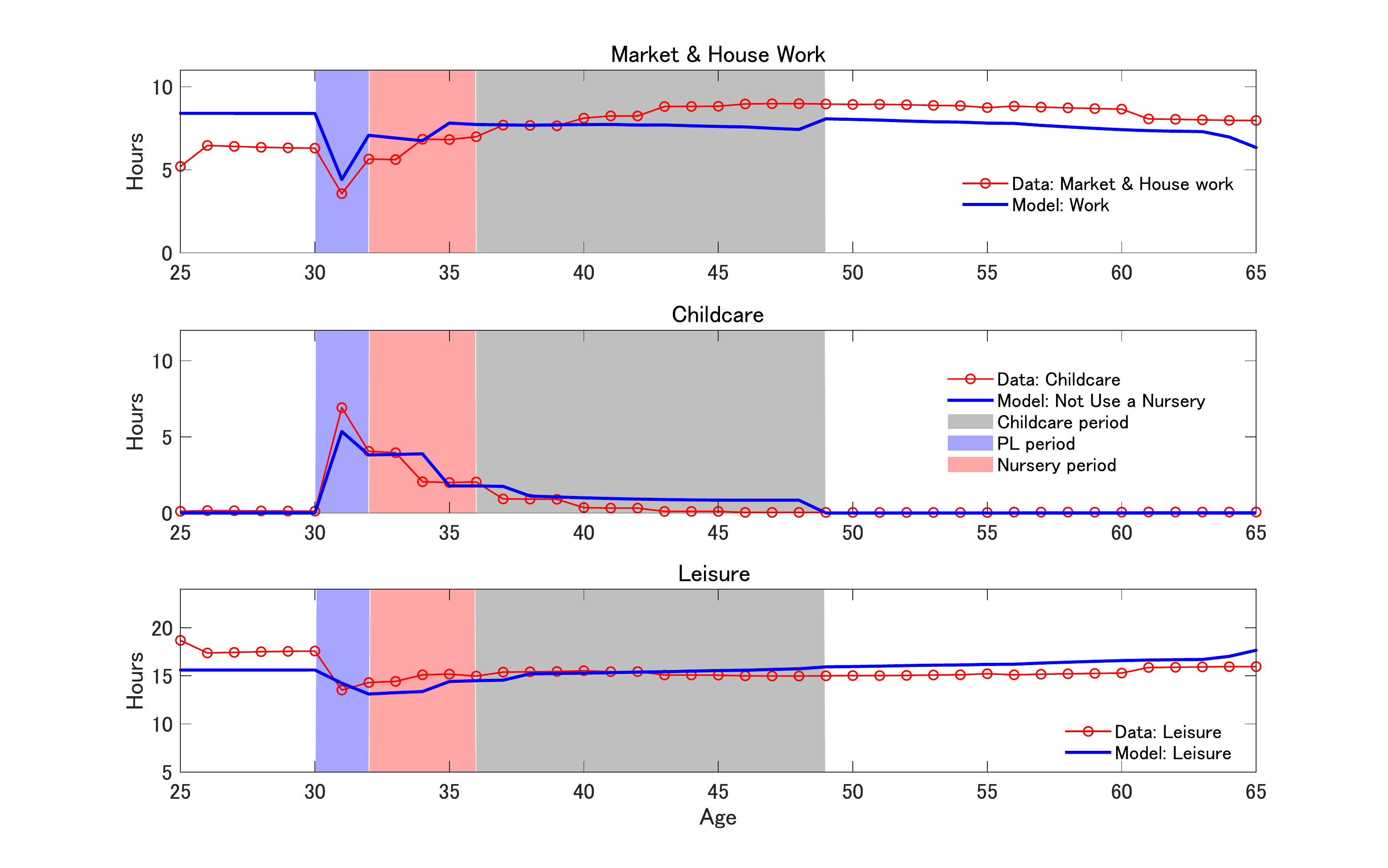}

\end{center}
\end{figure}

Figure \ref{fig:result_model_validation} 
presents the computed working, childcare, and leisure time from the model, compared with the aggregated data from the STULA, as explained in Section  \ref{sec:DATA} (indicated by the red line with markers). Panel (a) examines the situation of a married couple with a college education who use a nursery school, while panel (b) focuses on a married couple with a high school education and no nursery school use.

The first row of graphs in Panels (a) and (b) displays the data for the total hours, combining market work and housework (depicted by the solid red line with markers ``$\circ$''). The model-calculated working hours (solid blue line) align significantly with the data. The working hours calculated from the model are overestimated compared with the actual data during the period up to nursery (the light red shade) but are underestimated for ages thereafter (the light gray shade).

From the second row of graphs in Panel (a), it becomes apparent that, when a nursery is used, the model-estimated parenting time falls below the data (the red line with the circle marker) when the child is between 2 and 4 years old (the light red shade). In contrast, in Panel (b), which shows the case of nursery school, the childcare time predicted by the model exceeds the corresponding data throughout all the periods until the child reaches adulthood, except the PL period (the light blue shade).

\subsection{Model Validation for Earnings and the Child Penalty}

\begin{figure}
\caption{Comparison between the Model and the Data on the Child Penalty}
\label{fig:Child_Penalty}
\begin{center}

\includegraphics[width=14cm,height=8cm]{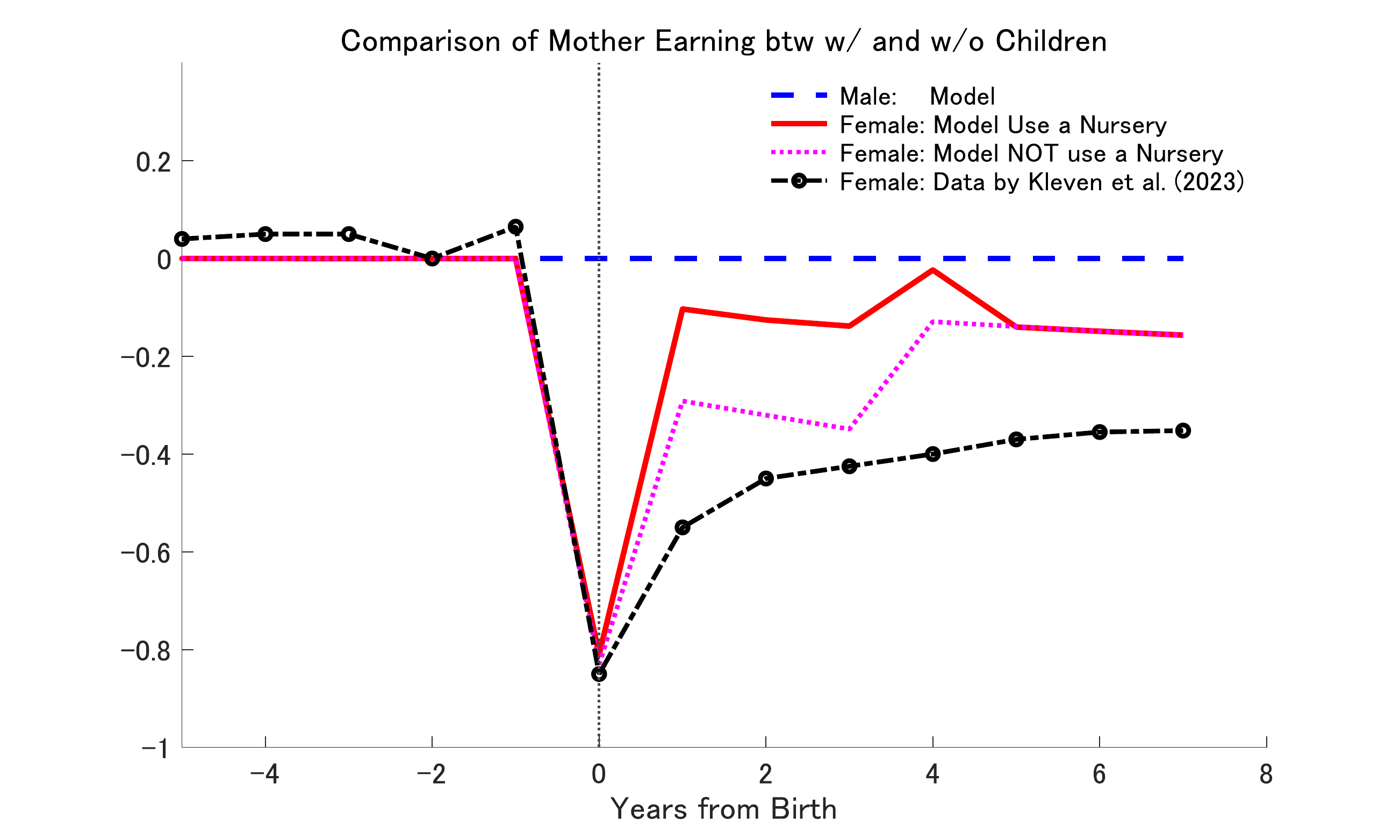}

\end{center}
\end{figure}

As explained in Section 1, the child penalty is one of the motivations for this study. To validate our life-cycle model, we employ our estimated parameters for Japanese women in Table  \ref{tab:estimate_parameters} to calculate, based on computations derived from our model, the rate at which the earnings levels of women who have given birth return to those of women who have not given birth. We compare our model's results with those of \cite{Kleven2023} on child penalties in Japan. As illustrated in Figure \ref{fig:Child_Penalty} 
the horizontal axis represents the years following childbirth. The empirical results, shown by the black dashed line with the ``$\circ$'' markers, depicts a labor earnings decline of about 85\% after childbirth. Seven years later, this value only partially recovers to around 40\%. In contrast, our model, even without considering nursery school, demonstrates an income reduction of only 35\% from the second year, followed by a swift recovery from the third year onward. The following potential factors may explain the notable difference between our model and the empirical evidence: first, women transitioning to lower-paying roles with greater working-hour flexibility due to career changes; second, wage structures reflecting employers' gender bias; and, finally, entrenched gender norms, as surveyed by \citet{cortes2023}.

In other words, when a couple chooses to raise children, it leads to a voluntary reduction in the wife's market working hours. This voluntary reduction in market working hours by the wife is thought to account for approximately half of the magnitude of the child penalty. The remaining half can be attributed to involuntary factors, as described above.

\section{Policy Simulations}
\label{sec:SIMULATION}
This section presents the findings from four types of counterfactual simulations conducted using the estimated parameters in Table  \ref{tab:estimate_parameters}, as depicted in Figures \ref{fig:simulation_current_policy} and \ref{fig:simulation_counterfactual}. 
First, as in Section \ref{sec:NUMEARICAL}, we assume a 50\% income RR for PL based on the current policies and compute the probability of childbearing and PL use, women's time allocation, and changes in labor income and household assets over the life cycle. In the first two simulations, we analyze the differences between (1) parents with a college education or higher and those with a high school education or lower and (2) parents with and without the use of childcare centers. Next, we conduct two counterfactual policy simulations related to alternative policies. The first policy simulation shows the change in households' life cycle decisions caused by an increase in the income RR for PL benefits, rising from 50\% to 75\%. The second simulation calculates the change in life cycle decisions of parents if their labor income is permanently increased by 10\%, as in  \citet{blundellChildrenTimeAllocation2018}.


\subsection{Effects of College Graduation }
\begin{figure}
\caption{Current Policy Evaluations}
\label{fig:simulation_current_policy}
\begin{center}

(a) Effects of College Graduation

\includegraphics[width=16cm,height=10cm]{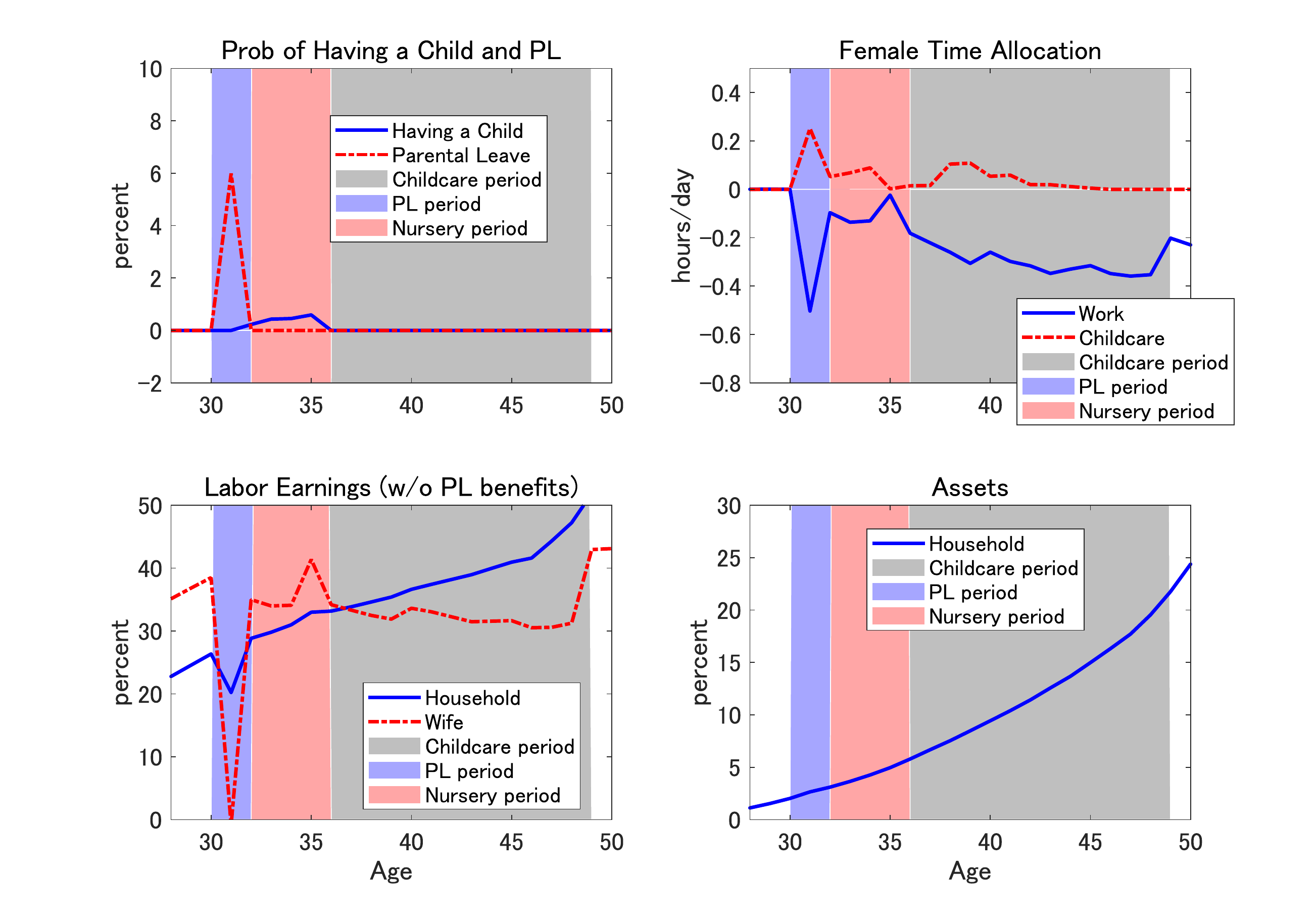}

\bigskip{}

(b) Effects of Using a Nursery

\includegraphics[width=16cm,height=10cm]{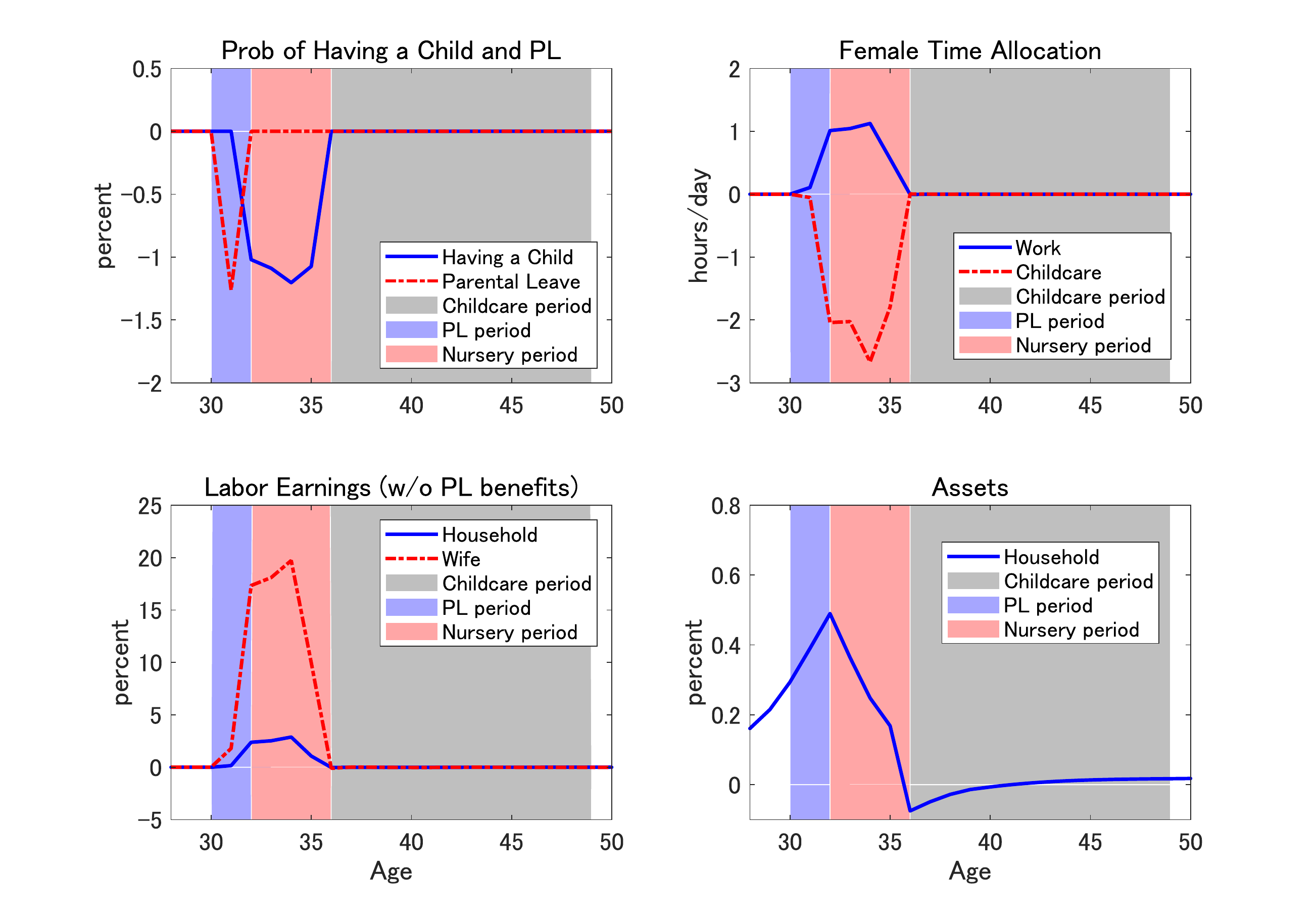}

\end{center}
\end{figure}

First, our analysis aims to investigate the impact of higher education. We achieve this by contrasting outcomes between parents with high school education and parents with college education over the life cycle. Panel (a) of Figure \ref{fig:simulation_current_policy} 
illustrates the differences between college graduates and high school graduates. As shown in Figure \ref{fig:data_LP_SR}, our model incorporates differences in education as differences in labor productivity. Essentially, households with college graduates have higher labor earnings over the life cycle due to the higher labor returns associated with a college degree.

The top left graph depicts variations in the probability of childbirth and taking PL. The probability of childbirth (represented by the blue solid line) is approximately 1 percentage point higher for college graduates than for high school graduates in their 30s. Furthermore, the probability of taking PL (indicated by the red dashed line) is about 6 percentage points greater for college graduates. 

We focus on the time allocation between work and childcare for mothers, as indicated by the top right graph. It becomes evident that college graduates increase their commitment to childcare (red dashed line) by roughly 15 minutes (0.2 hours) more per day compared with high school graduates upon the birth of their child. In contrast, they reduce their working hours (blue solid line) by approximately 30 minutes (0.5 hours) more per day than high school graduates.

The bottom left graph reveals disparities in the labor income profiles of husbands and wives. Regarding labor earnings, wives with a college degree experience an increase of over 30\% throughout their lives. Likewise, when both spouses within a marriage hold a college degree, there is a gradual rise from 20\% at age 30 to 50\% at age 50 in the whole household (the blue solid line). This pattern arises due to husbands' labor earnings being significantly influenced by the seniority system tied to their age. 
The bottom right graph illustrates variations in household asset profiles. The notable contrast in labor earnings based on educational attainments is reflected in household assets, which grow from 2.5\% before age 30 to over 15\% after age 45.

In summary, the key findings presented above can be described as follows: college graduates, who benefit from higher incomes, show a higher probability of childbearing and taking PL, resulting in temporary income losses compared with high school graduates. Moreover, college-educated women work fewer hours and devote more time to childcare.

It should be noted that our model treats individuals' choice of education as a given in this paper. Our numerical results suggest that, if women with homogeneous skills are allowed to choose their education, they are likely to choose to graduate from college. However, as analyzed by \cite{Greenwood2016}, if there is variability in agents' innate abilities, which introduces risk associated with college graduation and variability in the professional skills acquired at college, then variability in educational choices would also occur. Furthermore, there is a strong correlation between the educational attainment of spouses; our model assumes assortative mating by educational level, treating the spouse's education as a significant factor in marital decisions.

\subsection{Effects of Using a Nursery School}
As explained earlier, the use of a nursery school in Japan is determined less by the preferences of the couple than by the government's policy on the allocation of nursery schools.\footnote{This aspect was empirically analyzed in detail by \cite{Yamaguchi2018a}.}
Therefore, we assume that the choice of using nursery schools is given to the agents.\footnote{For further understanding of the secondary effects of nursery usage, one can refer to \cite{Yamaguchi2018b}'s study.}

We explore the impact of nursery school use in the context of college-educated parents. Panel (b) of Figure \ref{fig:simulation_current_policy} 
illustrates the contrast between the calibrated outcomes when nursery schools are utilized and when they are not. The graph demonstrates that, while the probability of childbirth decreases by around 1.0 percentage points, the probability of taking PL drops by around 1.2 percentage points. Notably, between the ages of 32 and 35, the introduction of nursery schools leads to a reduction of over 2 hours of childcare time per day, accompanied by a substantial increase in working hours.

Regarding labor earnings, the use of nursery schools increases the earnings of college-educated wives in their early 30s by almost 20 percentage points compared with not using nursery schools. Consequently, the household income increases by approximately 3 percentage points. In terms of household assets, the increased number of hours worked by parents when their child attends a nursery school contributes to a slightly greater asset accumulation.

It is important to note that the use of nursery schools exerts a significant impact on mothers' career choices and earnings progression, and it is expected to affect the household asset accumulation and consumption behavior in later life as well as pension benefits in retirement. Unfortunately, our model does not incorporate mothers' career choices. If we were to incorporate career choices and the resulting wage increases into the model, the changes over the life cycle could be even more significant. Indeed, focusing on these career choices is a crucial research topic. \cite{Goldin2014} and \cite{Goldin2021} highlighted the gender gap arising from differences in the choice of \textit{greedy work} across professions between men and women. While not a life-cycle model like ours, \cite{Erosa2022} analyzed numerically how women's career choices contribute to the gender gap, focusing on the different wage structures across occupations in the U.S. This model was adapted to the Japanese context by \cite{Yanagimoto2024}.

\subsection{Effects of the 25\% Rise in Income Replacement Rate (RR) for PL}
In the following, we conduct two counterfactual analyses that assume an increase in household income. The first simulation focuses on the changes in household behavior resulting from raising the PL benefit from 50\% to 75\% in terms of RR. This adjustment results in a 25-percentage point decrease in income loss during the PL period compared with the current program.

\begin{figure}
\caption{Counterfactual Policy Evaluations}
\label{fig:simulation_counterfactual}
\begin{center}

(a) Effects of a  25\% Rise in the Income Replacement Rate on PL

\includegraphics[width=16cm,height=10cm]{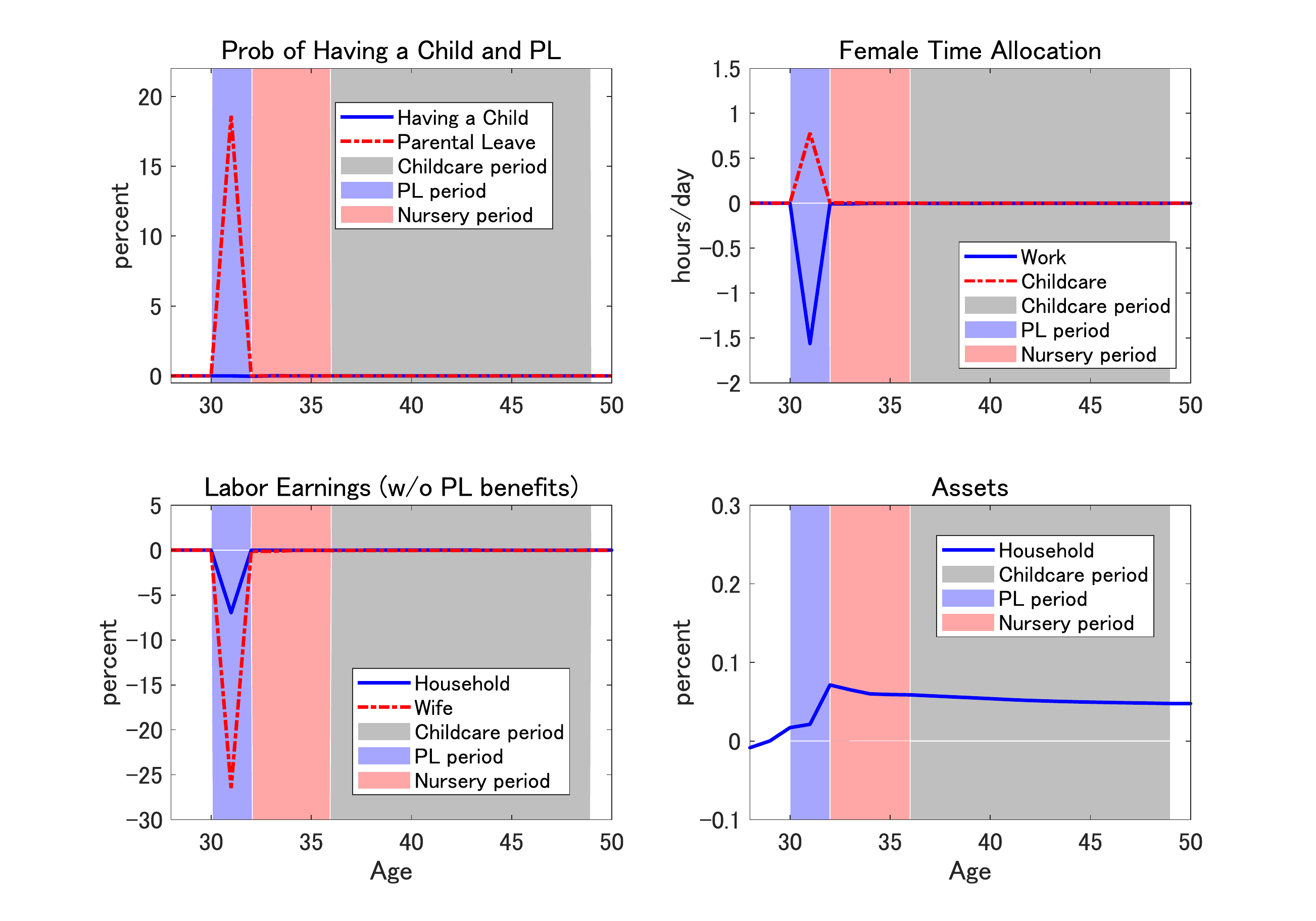}

\bigskip{}

(b) Effects of a 10\% Wage Increase

\includegraphics[width=16cm,height=10cm]{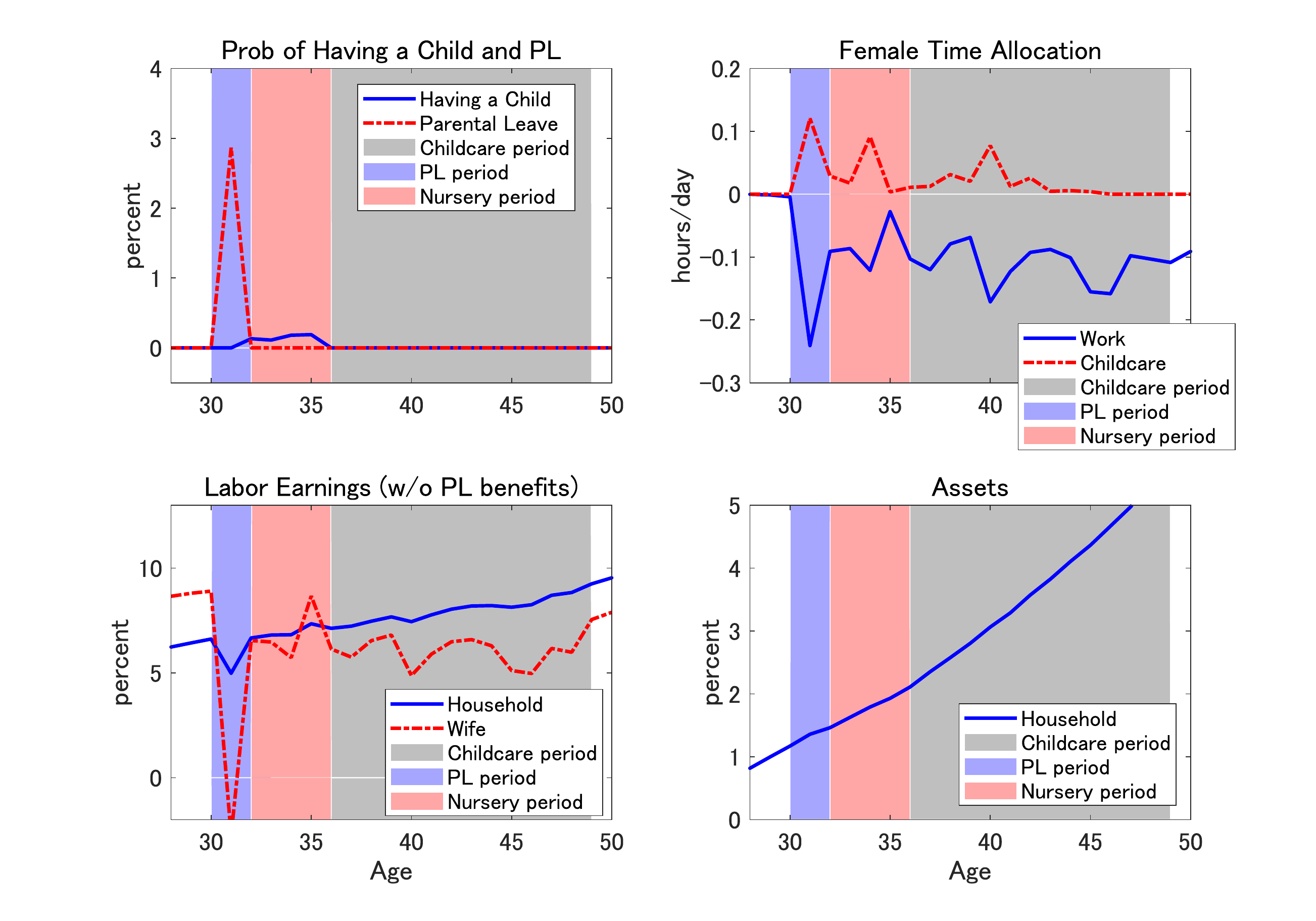}

\end{center}
\end{figure}

Panel (a) of Figure \ref{fig:simulation_counterfactual} 
illustrates the change in choices among people before and after the policy change. The top left graph shows that the probability of taking PL increases by approximately 20 percentage points due to the increase in PL benefit, while the probability of childbirth remains unchanged. Moreover, the rise in PL benefits and the resulting higher likelihood of taking PL contribute to a maximum increase of around 45 minutes (0.75 hours) in childcare time during parents' early 30s, accompanied by a reduction of about 1.5 hours in working time. Additionally, in line with the increased probability of taking PL, women's average earnings are lower by approximately 25 percentage points, leading to a decline of around 10 percentage points in household earnings within this period.

\subsection{Effects of a Permanent 10\% Wage Increase}

The second counterfactual examines the change in agents' behavior when the income of both spouses increases permanently by 10\% over the life cycle. Panel (b) of Figure \ref{fig:simulation_counterfactual} 
illustrates the contrast between the calibrated outcomes after the 10\% income increase and those before it. The top left graph shows that the probability of taking PL increases by around 3.0 percentage points due to the augmented PL benefit. At the same time, there is no significant change in the probability of childbirth. Focusing on the change in the wife's time allocation, childcare time experiences a maximum increase of 6 minutes (0.1 hours), while labor time slightly decreases by 12 minutes (0.2 hours). These results suggest that, as their income steadily rises, wives allocate less time to leisure activities and more time to childcare. The trajectory of household assets also follows the pattern of increased earnings.

\section{Conclusion}
\label{sec:CONCLUSION}
By extending the model of married couples' decision making developed by \citet{blundellChildrenTimeAllocation2018} to a life-cycle model that focuses on the impact of children's age from birth to adulthood, we analyzed the time allocation of couples, especially wives, at each life stage. To this end, we employed the GMM to estimate the life-cycle model with aggregated Japanese data such as the STULA. This dataset encompasses responses from approximately 200,000 participants. Additionally, we used the Basic Survey on Wage Structure to explore how income disparities related to educational attainment affect parental time allocation. Employing a model aligned with this dataset, we conducted policy simulations to assess quantitatively the impact of income benefit for PL on childcare time and the impact of nursery school use on parental working hours and leisure time.

Our estimated results provided the following insights into the drop in women's income associated with childbirth and the recovery of income levels after childbirth. Our outcome corresponds to the 85\% reduction in maternal earnings observed during childbirth, commonly referred to as the ``child penalty.'' However, deviations between our results and the empirical evidence become apparent in the subsequent recovery phase of maternal earnings. This inconsistency becomes more pronounced 3 years after childbirth, reaching a peak discrepancy of around 50\%. Thus, the speed of the decline in earnings, as indicated by the concept of the child penalty, seems to be related to mothers' reduced participation in the labor market, particularly in the period from birth to about 2 years, when they allocate a larger portion of their time to childcare. Conversely, the decline in maternal earnings beyond the third year after childbirth cannot be attributed solely to a reduction in working hours due to childcare responsibilities.

Compared with the high school education scenarios of our simulations, cases in which a couple has a college education increased women's use of PL by 6\%, accompanied by a reduction in working hours of about 30 minutes per day. In contrast, the use of nursery schools led to an average increase in women's working hours of about 1 hour due to a concurrent reduction in childcare hours. However, this also results in a decrease of around 1.2\% in women's use of PL. Notably, a 25\% increase in the income RR for PL is associated with a nearly 20\% increase in PL use, while a consistent 10\% increase in wages is associated with a 3.0\% increase. Another significant finding is that increased family assets are associated with reduced female working hours and increased childcare hours.




\vspace{1\baselineskip}
\renewcommand{\baselinestretch}{0.7} \normalsize
\bibliographystyle{econ} 

\begin{small}
\bibliography{references_lifecycle}

\end{small}

\end{document}